\documentclass[aps,prc,twoside,twocolumn,nofootinbib,10pt,floatfix]{revtex4}
\usepackage{amsmath,amssymb}
\usepackage{graphicx,bm}
\usepackage{slashed}
\usepackage{epstopdf}
\usepackage{ulem} 
\usepackage[usenames]{color}
\usepackage{float}
\usepackage{hyperref}
\usepackage{subfigure}
\usepackage{rotating}
\usepackage{color}
\usepackage{multirow}
\usepackage{dcolumn}
\usepackage{overpic}
\usepackage{booktabs}
\usepackage{makecell}
\usepackage{arydshln} 
\usepackage{diagbox}
\usepackage{tabularx}
\usepackage{array}
\newcolumntype{|}{!{\vline}}

\renewcommand\sout{\bgroup \color{red} \ULdepth=-.5ex \ULset}

\newsavebox{\tablebox}
\usepackage{adjustbox}
\begin{document}
\title{Doubly charmed hexaquarks in the diquark picture}
\author{Hong-Tao An$^{1}$}\email{anht@mail.tsinghua.edu.cn}
\author{Si-Qiang Luo$^{2,3,4,5}$}\email{luosq15@lzu.edu.cn}
\author{Xiang Liu$^{2,3,4,5}$}\email{xiangliu@lzu.edu.cn}
\affiliation{
$^1$Department of Physics, Tsinghua University, Beijing 100084, China \\
$^2$School of Physical Science and Technology, Lanzhou University, Lanzhou 730000, China\\
$^3$Lanzhou Center for Theoretical Physics, Key Laboratory of Theoretical Physics of Gansu Province,\\
Key Laboratory of Quantum Theory and Applications of MoE,\\
Gansu Provincial Research Center for Basic Disciplines of Quantum Physics, Lanzhou University, Lanzhou 730000, China\\
$^4$MoE Frontiers Science Center for Rare Isotopes, Lanzhou University, Lanzhou 730000, China\\
$^5$Research Center for Hadron and CSR Physics, Lanzhou University and Institute of Modern Physics of CAS, Lanzhou 730000, China
}

\date{\today}
\begin{abstract}
We investigate doubly charmed hexaquark states within the diquark picture, by employing the constituent quark model and the quark-interchange model as our theoretical frameworks. Using the Gaussian expansion method, we systematically study these states, with calculating various properties such as mass spectra, internal contributions of each Hamiltonian component, root-mean-square radii, and two-body strong decay widths. Our analysis of the mass spectra reveals no stable state in this system. Furthermore, the root-mean-square radii suggest that the doubly charmed hexaquark states exhibit a compact configuration. By examining the decay widths, we identify potentially detectable states and their primary decay channels within each subsystem. Despite the large decay phase space, we still find narrow states with total widths of less than 10 MeV. This study provides a theoretical foundation for understanding the structures and interactions of doubly charmed hexaquark states and offers valuable insights for future experimental searches.
\end{abstract}
\maketitle

\section{Introduction}\label{sec1}

In the field of hadron physics, the study of exotic states beyond conventional mesons (\(q\bar{q}\)) and baryons (\(qqq\)) has been a central research focus~\cite{Chen:2016qju,Guo:2017jvc,Liu:2019zoy,Hosaka:2016pey,Jaffe:2004ph}. A pivotal breakthrough occurred in 2003 with the Belle collaboration's discovery of the \(X(3872)\)~\cite{Belle:2003nnu}, ushering in an era of rapid progress. Subsequent experiments by Collaborations such as LHCb, BaBar, Belle, and BESIII have unveiled a spectrum of unconventional hadronic states, including the \(P_c/P_{cs}\) pentaquarks \cite{LHCb:2015yax,LHCb:2019kea,LHCb:2020jpq,LHCb:2022ogu} and charmonium-like \(XYZ\) states \cite{Belle:2004lle,LHCb:2022aki,LHCb:2021uow,BESIII:2016bnd,BESIII:2013ris,BESIII:2013qmu}. These discoveries have motivated extensive theoretical efforts to elucidate their internal structures, exploring configurations such as compact tetraquarks, molecular states, hybrids, and glueballs \cite{Chen:2022asf,Dong:2021juy,Brambilla:2019esw}.

Although the hidden-charm sector has dominated this landscape, recent observations of the doubly charmed baryon \(\Xi^{++}_{cc}(3621)\)~\cite{LHCb:2017iph,LHCb:2018pcs} and the tetraquark candidate \(T_{cc}^+(3875)\)~\cite{LHCb:2021vvq,LHCb:2021auc} by LHCb have shifted attention to the double-charm sector. This raises intriguing questions: Could doubly charmed pentaquarks exist, analogous to their hidden-charm counterparts? Moreover, what about doubly charmed hexaquarks? 
The latter, as a six-quark system, brings computational challenges due to the complexity of its inherent six-body problem. Current theoretical studies often adopt approximation methods.
For example, Liu {\it et al.} investigated the mass spectra of these systems using the chromomagnetic interaction (CMI) model \cite{Liu:2022rzu}. Complementary to this approach, the one-boson-exchange potential model has been employed to examine hadron-level interactions and predict the properties of doubly charmed hexaquarks within the framework of two-body molecular states, as demonstrated in Refs. \cite{Meguro:2011nr,Cheng:2022vgy,Shah:2024thr,Dong:2021bvy,Qi:2024dqz,Vijande:2016nzk,Andreev:2024orz}. And such similar triply-charmed hexaquarks, and the hadronic molecular states $D^*D^*\bar{D}^*/D^*D^*D^*$ are also studied with QCD sum rules~\cite{Wang:2020jqu,Wang:2020fuh}. Further insights have emerged from exploratory lattice QCD calculations, which address the binding characteristics of these systems \cite{Geng:2024dpk}. While these efforts provide valuable insights, the reliance on diverse approximations underscores the necessity for novel methodologies to resolve persistent theoretical ambiguities and deepen our understanding of hexaquark dynamics.

A promising framework for simplifying such systems is the diquark model ~\cite{Ida:1966ev,Anselmino:1992vg,Barabanov:2020jvn}, proven effective in hadron spectroscopy. 
For example, the diquark-heavy quark picture has been widely applied to single heavy baryon systems \cite{Ebert:2011kk,Chen:2016iyi,Chen:2017gnu,Chen:2018orb}, where mass spectra could be well depicted.
In Ref.~\cite{Chen:2021eyk}, relevant results further implied that diquark picture highly matches the symmetry of spectra for the singly charmed baryons. 
Thus, the diquark is an effective approach for researching baryon spectra.
As an important continuation of singly charmed baryon spectroscopy, the hexaquark states in this work also contain several light-flavor quarks. 
With the experiences in baryon systems, we propose that hexaquark spectra also have similar symmetry with the same approach.
Based on this, we further expand and apply the diquark-diquark-heavy quark-heavy quark picture 
(a four-body system) to the doubly charmed hexaquark state system (a six-body system).
By ``freezing" the diquarks’ internal degrees of freedom, this simplification reduces computational complexity while preserving key physical features, akin to charm baryon treatments.
Furthermore, the model also demonstrates unique value in studying multi-quark states with other configurations— for instance, in the compact diquark-antidiquark, diquark+diquark+antiquark, and diquark-diquark-diquark configurations, it offers an effective way to investigate hidden-charm and hidden-bottom $XYZ$ exotic states \cite{Zhu:2016arf,Shi:2021jyr,Bicudo:2015vta}  (like \(P_{c}\), \(P_{cs}\) states \cite{Ali:2019clg,Lebed:2015tna}).
In addition, R. Lebed et al. have successfully applied the dynamical diquark model \cite{Brodsky:2014xia,Lebed:2017min,Lebed:2025xbz}, which explicitly treats the relative motion of color sources, allowing a more realistic
description of spatial separation and interaction dynamics, to multiple flavor sectors: $s\bar{s}q\bar{q}$ \cite{Jafarzade:2025qvx},
$c\bar{c}q\bar{q}$ \cite{Giron:2019bcs,Giron:2019cfc,Giron:2020fvd}, $c\bar{c}c\bar{c}$ \cite{Giron:2020wpx}, $cc\bar{q}\bar{q}$ \cite{Mutuk:2024vzv}, and $c\bar{c}qqq$ \cite{Giron:2021fnl}.

Notably, this framework suggests a structural parallel between the \(T_{cc}^+(3875)\) tetraquark (composed of two charm quarks and two light antiquarks) and a hypothetical doubly charmed hexaquark.
Replacing two light antiquarks $\bar{q}\bar{q}$ in \(T_{cc}^+(3875)\) with two light diquarks $[qq][qq]$ would yield a hexaquark configuration (\(ccqqqq\)), potentially mirroring properties of its tetraquark counterpart. 
Establishing such connections theoretically reveals universal features of multiquark systems, including potential shared properties like binding mechanisms and decay modes. Experimentally, it provides guidance for detecting elusive hexaquarks by leveraging known tetraquark signatures.

In this work, we systematically investigate the doubly charmed hexaquark system \(ccqqqq\) ($q = n, ~s; ~n = u, ~d$) using the Gaussian expansion method within the framework of the constituent quark model.
Our calculations include: the mass spectra for all flavor configurations, corresponding internal mass contributions, root-mean-square (RMS) radii, and partial (total) decay widths. 

The paper is organized as follows. After the Introduction, Section \ref{sec2} details the theoretical framework, including the effective Hamiltonian, hexaquark configuration, and the calculation methods for root-mean-square radii and two-body strong decays. Then in Section \ref{sec3}, we present results for mass spectra, internal structure, and decay properties. Finally, the key findings and implications for future studies are summarized in Section \ref{sec4}.

\section{Theoretical framework for doubly charmed hexaquark system}\label{sec2}
\subsection{The effective Hamiltonian
}\label{sec21}

\begin{table*}[t]
\caption{Parameters of the Hamiltonian determined by fitting the singly-charmed baryon masses.
The $M_{\rm the}$, $M_{\rm exp}$, and Error are the theoretical value, the experimental value, and the error between them, respectively.
}\label{para}
\renewcommand\arraystretch{1.8}
\renewcommand\tabcolsep{2.8pt}
\begin{tabular}{p{2.2cm}|m{2.5cm}|m{2.5cm}|m{2.5cm}|m{2.5cm}|m{2.5cm}|m{2.5cm}|m{2.5cm}|m{2.5cm}}
\toprule[1.50pt]
\toprule[0.50pt]
Parameter&\multicolumn{1}{c|}{$m_{n}$}&\multicolumn{1}{c|}{$m_{c}$}&\multicolumn{2}{c|}{$\kappa$}&\multicolumn{2}{c|}{$a_{0}$}&\multicolumn{2}{c}{$\kappa_{0}$}\\
\Xcline{1-9}{0.5pt}
Value&\multicolumn{1}{c|}{321.0 MeV}&\multicolumn{1}{c|}{1508.0 MeV}&\multicolumn{2}{c|}{$1.5\times10^{2}$ MeV fm}&\multicolumn{2}{c|}{$2.0\times10^{-2}$ $\rm (MeV^{-1}fm)^{1/2}$}&\multicolumn{2}{c}{$2.8\times10^{2}$ MeV}\\
\Xcline{1-9}{0.5pt}
Parameter&\multicolumn{1}{c|}{$m_{s}$}&\multicolumn{1}{c|}{$D$}&\multicolumn{2}{c|}{$\alpha$}&\multicolumn{2}{c|}{$\beta$}&\multicolumn{2}{c}{$\gamma$}\\
\Xcline{1-9}{0.5pt}
Value&\multicolumn{1}{c|}{642.0 MeV\quad}& \multicolumn{1}{c|}{1033.0 MeV}&\multicolumn{2}{c|}{1.3 $\rm fm^{-1}$}& \multicolumn{2}{c|}{$-4.4\times10^{-4}$ $\rm (MeV fm)^{-1}$}&\multicolumn{2}{c}{$-5.6\times10^{-4}$ $\rm MeV^{-1}$}\\
\midrule[1.5pt]
Baryon&\multicolumn{1}{c|}{$\Lambda_{c}$}&\multicolumn{1}{c|}{$\Sigma_{c}$}&\multicolumn{1}{c|}{$\Sigma^{*}_{c}$}&\multicolumn{1}{c|}{$\Xi_{c}$}&
\multicolumn{1}{c|}{$\Xi'_{c}$}&\multicolumn{1}{c|}{$\Xi^{*}_{c}$}&\multicolumn{1}{c|}{$\Omega_{c}$}&\multicolumn{1}{c}{$\Omega^{*}_{c}$}
\\ 
\midrule[0.5pt]
$M_{\rm the}$ (MeV)&\multicolumn{1}{c|}{2279.8}&\multicolumn{1}{c|}{2447.2}&
\multicolumn{1}{c|}{2534.2}&\multicolumn{1}{c|}{2487.2}&\multicolumn{1}{c|}{2572.4}&\multicolumn{1}{c|}{2648.7}&\multicolumn{1}{c|}{2678.1}&\multicolumn{1}{c}{2747.2}
\\ \midrule[0.5pt]
$M_{\rm exp}$ (MeV)&\multicolumn{1}{c|}{2286.5}&\multicolumn{1}{c|}{2452.9}&\multicolumn{1}{c|}{2517.5\quad}&\multicolumn{1}{c|}{2467.8}&\multicolumn{1}{c|}{2577.4}&\multicolumn{1}{c|}{2645.9}&\multicolumn{1}{c|}{2695.2}& \multicolumn{1}{c}{2765.9}
\\ \midrule[0.5pt]
Error (MeV)&\multicolumn{1}{c|}{-6.7}&  \multicolumn{1}{c|}{-6.8\quad}& \multicolumn{1}{p{1.2cm}|}{\quad 15.8}&\multicolumn{1}{c|}{19.3}&\multicolumn{1}{p{1.6cm}|}{\quad\quad 5.0 }&\multicolumn{1}{c|}{ 3.2 }&\multicolumn{1}{p{1.3cm}|}{\quad -17.1}&\multicolumn{1}{c}{ -18.7 }\\
\bottomrule[0.50pt]
\bottomrule[1.50pt]
\end{tabular}
\end{table*}

In the constituent quark model, the nonrelativistic Hamiltonian incorporates three key components to describe quark interactions and calculate the properties of ground-state hadrons: a linear confinement potential (modeling long-range quark confinement), a Coulomb-like potential (accounting for short-range chromoelectric interactions), and a hyperfine interaction potential (arising from spin-dependent forces). The Hamiltonian is thus expressed as:
\begin{eqnarray}\label{Eq1}
H=\sum_{i=1}^{4}(m_{i}+T_{i})-T_{\rm cm}-\frac{3}{4}\sum_{i<j}^{4}
\frac{\lambda^{c}_{i}}{2}.\frac{\lambda^{c}_{j}}{2}(V^{\rm Con}_{ij}+V^{\rm SS}_{ij}).\nonumber\\
\end{eqnarray}
Here, $m_{i}$ represents the mass of the $i$-th constituent quark; $T_{i}=\textbf{p}^{2}_{i}/(2m_{i})$ stands for the kinetic energy of the $i$-th quark; 
$T_{\rm cm}$ denotes the center-of-mass kinetic energy of the corresponding hexaquark system;
and $\lambda^{c}_{i}$ is the $SU(3)$ color operator associated with the $i$-th quark.

The confinement potential $V^{{\rm Con}}_{ij}$ and spin-spin interaction potential $V_{ij}^{{\rm SS}}$ are defined as
\begin{eqnarray}\label{Eq2}
V^{{\rm Con}}_{ij}&=&-\frac{\kappa}{r_{ij}}+\frac{r_{ij}}{a^{2}_{0}}-D,\nonumber\\
V^{{\rm SS}}_{ij}&=&\frac{\kappa'}{m_{i}m_{j}}\frac{1}{r_{0ij}r_{ij}}e^{-r^{2}_{ij}/r^{2}_{0ij}}\sigma_{i}\cdot\sigma_{j},
\end{eqnarray}
where $r_{ij}=|\textbf{r}_{i}-\textbf{r}_{j}|$ is the interquark distance between the $i$-th and the $j$-th quarks, and $\sigma_{i}$ denotes the $SU(2)$ spin operator of the $i$-th quark.
The parameters $r_{0ij}$ and $\kappa'$ incorporate explicit 
mass dependence
\begin{eqnarray}\label{Eq3}
r_{0ij}&=&1/(\alpha+\beta\frac{m_{i} m_{j}}{m_{i}+m_{j}}),\nonumber\\
\kappa'&=&\kappa_{0}(1+\gamma\frac{m_{i} m_{j}}{m_{i}+m_{j}}).
\end{eqnarray}

The numerical values of the parameters in Eqs. (\ref{Eq2})-(\ref{Eq3}) are listed in Table \ref{para}.
For completeness, the table also includes the calculated mass of singly-charmed baryons within the heavy-quark-diquark configuration, alongside their experimental values for comparative analysis.

\subsection{Hexaquark configuration
}\label{sec22}

When two light quarks are treated as a tightly bound diquark, 
the doubly charmed hexaquark system reduces to a four-body system in the [diquark-diquark]-[quark-quark] configuration.

The total wave function comprises flavor, spatial, color, and spin components. 
In the flavor structure, six distinct flavor configurations exist for the doubly charmed hexaquark system: $[nn][nn]cc$, $[ss][ss]cc$, $[ns][ns]cc$, $[nn][ns]cc$, $[nn][ss]cc$, and $[ns][ss]cc$.

In the color structure,  
a diquark in the color-antitriplet representation ($(qq)^{\bar{3}_{c}}$) is considered “good” due to its attractive confinement potential, while
the color-sextet $(qq)^{6_{c}}$ (“bad” diquark) exhibits repulsive interactions.  Only color-singlet configurations are physically admissible. The color decomposition proceeds as:
\begin{eqnarray}\label{color}
&&([3\otimes3]\otimes[3\otimes3])\otimes(3\otimes3)\nonumber\\
&=&([\bar{3}\oplus6]\otimes[\bar{3}\oplus6])\otimes(\bar{3}\oplus6)\nonumber\to(\bar{3}\otimes\bar{3})\otimes(\bar{3}\oplus6)\\
&=&(3\oplus\bar{6})\otimes(\bar{3}\oplus6)
=(3\otimes\bar{3})\oplus(\bar{6}\otimes6)\oplus\cdots
\end{eqnarray}
From Eq.~(\ref{color}), two color-singlet configurations emerge:
\begin{eqnarray}\label{color1}
\Psi_{1}&=&|[(qq)^{\bar{3}_{c}}(qq)^{\bar{3}_{c}}]^{3_{c}}(cc)^{\bar{3}_{c}}\rangle, \nonumber\\
\Psi_{2}&=&|[(qq)^{\bar{3}_{c}}(qq)^{\bar{3}_{c}}]^{\bar{6}_{c}}(cc)^{6_{c}}\rangle.
\end{eqnarray}
In the notation $|[(q_{1}q_{2})^{\bar{3}_{c}}(q_{3}q_{4})^{\bar{3}_{c}}]^{\rm color_{1}}(cc)^{\rm color_{2}}\rangle$,
$\rm color_{1}$ and $\rm color_{2}$ stand for the color representations of diquark pair and charm-quark pair, respectively.

In the spin structure, the system admits 20 spin configurations. The general spin wave function is expressed as 
$|[(q_{1}q_{2})_{\rm spin_{1}}(q_{3}q_{4})_{\rm spin_{2}}]_{\rm spin_{3}}(cc)_{\rm spin_{4}}\rangle_{\rm spin_{5}}$, where
$\rm spin_{1}$ and $\rm spin_{2}$ represent the spins of the light diquarks ($q_{1}q_{2}$) and ($q_{3}q_{4}$), respectively.
While, $\rm spin_{3}$ and $\rm spin_{4}$ are the total spin of charm-quark pair ($cc$) 
and diquark pair $[q_{1}q_{2}][q_{3}q_{4}]$, respectively.
Finally, $\rm spin_{5}$ represents the total spin of the doubly charmed hexaquark state. All possible spin wave functions are tabulated in Table~\ref{spin}.

Based on the symmetrized configurations of flavor, color, and spin spaces, we systematically construct the combined flavor$\otimes$color$\otimes$spin space for the doubly charmed hexaquark system. For the ground state, the spatial wave function is symmetric under the exchange of any two identical quarks or diquarks. In accordance with the Spin-Statistics Theorem, this necessitates that the 
flavor$\otimes$color$\otimes$spin component must be antisymmetric (symmetric) under the permutation of identical quarks (diquarks). 

This framework parallels earlier theoretical studies of the $\Theta^{+}$ pentaquark \cite{LEPS:2003wug}, where Jaffe and Wilczek \cite{Jaffe:2003sg} treated two diquarks as bosonic point-like constituents within a diquark-diquark-antiquark configuration. 
Their model predicted the $\Theta^{+}$ as a bound state of two identical diquarks and an anti-strange quark, alongside an isospin 
$I = 3/2$ multiplet including $\Xi^{--}_{5}$ ($S = -2$, $J^{P} = 1/2^{+}$) near 1750 MeV \cite{Jaffe:2003sg}. 
Subsequent works by Liu \textit{et al.} extended this approach to compute magnetic moments of analogous pentaquark states \cite{Li:2003cb,Liu:2003ab,Huang:2004tn,Zhang:2004xt}. 
While experimental confirmation of $\Theta^{+}$ and $\Xi^{--}_{5}$ remains elusive \cite{BES:2004kia,Hicks:2012zz}, the symmetry-driven methodology developed for these systems provides a valuable foundation for constructing hexaquark wave functions. 

Guided by these principles, we derive total wave functions for distinct flavor configurations that rigorously satisfy the Spin-Statistics Theorem. The resulting configurations are tabulated in Tables \ref{ccnnnn}--\ref{ccnnss}.  

A critical distinction arises in the role of interaction terms: while Coulomb and linear confinement potentials do not induce color-spin mixing, the hyperfine interaction term does. The hyperfine interaction strength scales as $1/m_i m_j$, where $m_i$ and $m_j$ denote the masses of the charm quark or diquarks ($[nn]$, $[ns]$, $[ss]$). Given that all constituent masses exceed 1 GeV, color-spin mixing is strongly suppressed in the doubly charmed hexaquark system. This suppression significantly simplifies the spectroscopic analysis compared to light-quark systems.

\begin{table*}[t]
\centering
\caption{All possible spin wave functions in the [diquark–diquark]-[quark-quark] configuration.}\label{spin}
\renewcommand\arraystretch{1.5}
\renewcommand\tabcolsep{2.5pt}
\begin{tabular*}{\textwidth}{@{\extracolsep{\fill}}ccccc}
\toprule[1.50pt]
\toprule[0.50pt]
\multirow{1}*{$J=3$}
&\multicolumn{1}{r}{$\chi_{1}=|[(qq)_{1}(qq)_{1}]_{2}[cc]_{1}\rangle_{3}$}\\
\midrule[0.5pt]
\multirow{2}*{$J=2$}
&\multicolumn{1}{r}{$\chi_{2}=|[(qq)_{1}(qq)_{1}]_{2}[cc]_{1}\rangle_{2}$}
&\multicolumn{1}{r}{$\chi_{3}=|[(qq)_{1}(qq)_{1}]_{2}[cc]_{0}\rangle_{2}$}
&\multicolumn{1}{r}{$\chi_{4}=|[(qq)_{1}(qq)_{1}]_{1}[cc]_{1}\rangle_{2}$}
&\multicolumn{1}{r}{$\chi_{5}=|[(qq)_{1}(qq)_{0}]_{1}[cc]_{1}\rangle_{2}$}\\
&\multicolumn{1}{r}{$\chi_{6}=|[(qq)_{0}(qq)_{1}]_{1}[cc]_{1}\rangle_{2}$}
\\
\midrule[0.5pt]
\multirow{3}*{$J=1$}
&\multicolumn{1}{r}{$\chi_{7}=|[(qq)_{1}(qq)_{1}]_{2}[cc]_{1}\rangle_{1}$}
&\multicolumn{1}{r}{$\chi_{8}=|[(qq)_{1}(qq)_{1}]_{1}[cc]_{1}\rangle_{1}$}
&\multicolumn{1}{r}{$\chi_{9}=|[(qq)_{1}(qq)_{1}]_{0}[cc]_{1}\rangle_{1}$}
&\multicolumn{1}{r}{$\chi_{10}=|[(qq)_{1}(qq)_{1}]_{1}[cc]_{0}\rangle_{1}$}\\
&\multicolumn{1}{r}{$\chi_{11}=|[(qq)_{1}(qq)_{0}]_{1}[cc]_{1}\rangle_{1}$}
&\multicolumn{1}{r}{$\chi_{12}=|[(qq)_{0}(qq)_{1}]_{1}[cc]_{1}\rangle_{1}$}
&\multicolumn{1}{r}{$\chi_{13}=|[(qq)_{1}(qq)_{0}]_{1}[cc]_{0}\rangle_{1}$}
&\multicolumn{1}{r}{$\chi_{14}=|[(qq)_{0}(qq)_{1}]_{1}[cc]_{0}\rangle_{1}$}\\
&\multicolumn{1}{r}{$\chi_{15}=|[(qq)_{0}(qq)_{0}]_{0}[cc]_{1}\rangle_{1}$}
\\
\midrule[0.5pt]
\multirow{2}*{$J=0$}
&\multicolumn{1}{r}{$\chi_{16}=|[(qq)_{1}(qq)_{1}]_{1}[cc]_{1}\rangle_{0}$}
&\multicolumn{1}{r}{$\chi_{17}=|[(qq)_{1}(qq)_{0}]_{1}[cc]_{1}\rangle_{0}$}
&\multicolumn{1}{r}{$\chi_{18}=|[(qq)_{0}(qq)_{1}]_{1}[cc]_{1}\rangle_{0}$}
&\multicolumn{1}{r}{$\chi_{19}=|[(qq)_{1}(qq)_{1}]_{0}[cc]_{0}\rangle_{0}$}\\
&\multicolumn{1}{r}{$\chi_{20}=|[(qq)_{0}(qq)_{0}]_{0}[cc]_{0}\rangle_{0}$}
\\
\bottomrule[0.50pt]
\bottomrule[1.50pt]
\end{tabular*}
\end{table*}

\subsection{Numerical calculation method}\label{sec23}
\subsubsection{Gaussian expansion method}\label{sec231}


In the spatial space, the Jacobi coordinates for a four-body system are defined in terms of single-particle coordinates $\textbf{x}_{i}$ $(i=1,2,3,4)$ as follows:
\begin{eqnarray}\label{jacobi}
&&\xi_{1}=\sqrt{1/2}(\textbf{x}_{1}-\textbf{x}_{2}),\nonumber\\
&&\xi_{2}=\sqrt{1/2}(\textbf{x}_{3}-\textbf{x}_{4}),\nonumber\\
&&\xi_{3}=(\frac{m_{1}\textbf{x}_{1}+m_{2}\textbf{x}_{2}}{m_{1}+m_{2}})-(\frac{m_{3}\textbf{x}_{3}+m_{4}\textbf{x}_{4}}{m_{3}+m_{4}}),\\
&&\textbf{R}=\frac{m_{1}\textbf{x}_{1}+m_{2}\textbf{x}_{2}+m_{3}\textbf{x}_{3}+m_{4}\textbf{x}_{4}}{m_{1}+m_{2}+m_{3}+m_{4}}.\nonumber
\end{eqnarray}
Here, $\xi_{1}$ represents the relative Jacobi coordinate within the diquarks $[q_{1}q_{2}]$ and $[q_{3}q_{4}]$. 
Meanwhile, $\xi_{2}$ denotes the relative Jacobi coordinate between the charm quarks $c$.
While $\xi_{3}$ describes the relative Jacobi coordinate between the centers of mass
of the two diquarks $([qq][qq])$ and the two charm quarks $(cc)$.
Using above Jacobi coordinates (Eq.~(\ref{jacobi})), the spatial wave function with well-defined symmetry for the pairs (12) and (34) can be readily constructed.
In the center-of-mass frame of the four-body system ($\textbf{R}=0$) and
the number of Jacobi coordinates reduces to three (see Fig. \ref{fig2}).

The kinetic term in the Hamiltonian (Eq. (\ref{Eq1})) can be
simplified for calculations. 
Denoted as $T_{c}$, it is expressed as:
\begin{eqnarray}\label{kinetic term}
T_{c}=\sum^{4}_{i=1}\frac{\textbf{p}^{2}_{x_{i}}}{2m_{i}}- \frac{\textbf{p}^{2}_{R}}{2M}= \frac{\textbf{p}^{2}_{\xi_{1}}}{2m'_{1}}+\frac{\textbf{p}^{2}_{\xi_{2}}}{2m'_{2}}+\frac{\textbf{p}^{2}_{\xi_{3}}}{2m'_{3}},
\end{eqnarray}
where the reduced masses $m'_{i}$ are given by $m'_{1}=\frac{2\times (m_{1}m_{2})}{m_{1}+m_{2}}$, $m'_{2}=\frac{2\times (m_{3}m_{4})}{m_{3}+m_{4}}$, and $m'_{3}=\frac{ (m_{1}+m_{2}) \times (m_{3}+m_{4})}{m_{1}+m_{2}+m_{3}+m_{4}}$.

The Gaussian expansion method (GEM) \cite{Hiyama:2003cu,Hiyama:2012sma,Brink:1998as} is employed to solve the four-body Schr\"{o}dinger equation.
This method has been widely applied to baryons \cite{Luo:2023sra}, tetraquarks \cite{Wu:2021rrc}, pentaquarks \cite{Yan:2023iie}, and few-body molecular states \cite{Wu:2021kbu,Luo:2021ggs,Luo:2022cun}.
The spatial wave function is expanded using a set of correlated Gaussian bases constructed from the Jacobi coordinates (Eq. (\ref{jacobi})):
\begin{eqnarray}\label{spatial}
&&\Psi(\xi_{1},\xi_{2},\xi_{3}) \nonumber\\
&=&\sum^{n_{1max}}_{n_{1}=1}\sum^{n_{2max}}_{n_{2}=1}\sum^{n_{3max}}_{n_{3}=1}
c_{n_{1}n_{2}n_{3}}\Psi^{n_{1}n_{2}n_{3}}(\xi_{1},\xi_{2},\xi_{3})
\nonumber\\
&=&\sum^{n_{1max}}_{n_{1}=1}\sum^{n_{2max}}_{n_{2}=1}\sum^{n_{3max}}_{n_{3}=1}
c_{n_{1}n_{2}n_{3}}\rm Exp[-v_{n_{1}}\xi^{2}_{1}-v_{n_{2}}\xi^{2}_{2}-v_{n_{3}}\xi^{2}_{3}].\nonumber\\
\end{eqnarray}
Here, $c_{n_{1}n_{2}n_{3}}$ are expansion coefficients determined by the Rayleigh-Ritz variational method, and $v_{n_{1}}$, $v_{n_{2}}$, $v_{n_{3}}$ are Gaussian range parameters chosen via a geometric progression:
\begin{eqnarray}\label{spatial1}
&&v_{n_{i}}=\frac{1}{r^{2}_{n_{i}}}, \quad \quad \quad   \quad r_{n_{i}}=r_{min_{i}}a^{n_{i}-1}, \nonumber\\ 
&&a=(\frac{r_{max_{i}}}{r_{min_{i}}})^{\frac{1}{n_{max_{i}}-1}} \quad \quad (i=1,2,3),
\end{eqnarray}
where $n_{max_{i}}$ is the number of Gaussian functions, and $a$ is the ratio coefficient.
There are three parameters \{$r_{max_{i}}$, $r_{min_{i}}$, $n_{max_{i}}$\} to be determined through the variation method. Stable results are achieved with \{5 fm, 0.7 fm, 5\}.

After the above preparations,  the eigenvalues of doubly charmed hexaquark system are obtained by solving the four-body Schr\"{o}dinger equation:
\begin{eqnarray}\label{equa}
\hat{H}\Psi_{\rm total}(\xi_{1},\xi_{2},\xi_{3})=E\Psi_{\rm total}(\xi_{1},\xi_{2},\xi_{3}),
\end{eqnarray}
where the Hamiltonian $\hat{H}$ (Eq.~(\ref{Eq1})) includes the kinetic term $T_{c}$ (Eq.~(\ref{kinetic term})) and the two-body interaction term between (di)quarks.
The total wave function $\Psi_{\rm total}(\xi_{1},\xi_{2},\xi_{3})$ combines the spatial part  (Eq.~(\ref{spatial})) and the flavor-color-spin part.

The normalization, kinetic, and potential matrix elements are calculated as
\begin{eqnarray}
\label{acca}
N^{nn'}&=&\langle\Psi^{n_{1}n_{2}n_{3}}(\xi_{1},\xi_{2},\xi_{3})\psi_{cs}|\psi'_{cs}\Psi^{n'_{1}n'_{2}n'_{3}}(\xi_{1},\xi_{2},\xi_{3})\rangle,\nonumber\\
V^{nn'}_\alpha&=&
\langle\Psi^{n_{1}n_{2}n_{3}}(\xi_{1},\xi_{2},\xi_{3})\psi_{cs}|
V_\alpha|\psi'_{cs}\Psi^{n'_{1}n'_{2}n'_{3}}(\xi_{1},\xi_{2},\xi_{3})\rangle ,\nonumber\\
T_{c}^{nn'}&=&\langle\Psi^{n_{1}n_{2}n_{3}}(\xi_{1},\xi_{2},\xi_{3})\psi_{cs}|T_{c}|\psi'_{cs}\Psi^{n'_{1}n'_{2}n'_{3}}(\xi_{1},\xi_{2},\xi_{3})\rangle,\nonumber\\
\end{eqnarray}
where $\psi_{cs}$ is the spin-color wave function, 
$V$ represents $V^{\rm Con}$ and $V^{\rm SS}$ in Eq.~(\ref{Eq2}), 
and $n$ simply refers to $\{n_{1}, n_{2}, n_{3}\}$ in Eq.~(\ref{spatial}). The $V_\alpha$ ($\alpha=$1-6) imply $V(r_{12})$, $V(r_{13})$, $V(r_{14})$, $V(r_{23})$, $V(r_{24})$, and $V(r_{34})$.
According to Eq.~(\ref{acca}), the Schr\"{o}dinger equation (Eq.~(\ref{equa})) is transformed into a generalized matrix eigenvalue problem:
\begin{eqnarray} \label{tven}
[T^{nn'}_{c}+\sum^{6}_{\alpha=1}V^{nn'}_{\alpha}-EN^{nn'}]C_{nn'}=0.
\end{eqnarray}
Solving this yields the eigenvalue $E$ and the internal mass contributions, as shown in Tables \ref{ccnnnn}-\ref{ccnnss}.

\subsubsection{Root-mean-square radii}\label{sec232}

To further probe the inner structure of the doubly charmed hexaquark, we calculate the root-mean-square (RMS) radii between all pairs of (di)quarks. 
This parameter is crucial for distinguishing between compact multiquark states and hadronic molecular states. Specifically, a molecular state typically exhibits minimal spatial overlap between hadronic constituents, whereas a compact multiquark state shows significant overlap \cite{Luo:2022cun}.

The  RMS radius $R_{ij}$ is defined as:
\begin{eqnarray}\label{rms}
R^{2}_{ij}=\int (\textbf{x}_{i}-\textbf{x}_{j})^{2}|\Psi(\xi_{1},\xi_{2},\xi_{3})|^{2}d\xi_{1}d\xi_{2}d\xi_{3}.
\end{eqnarray}
The results for the different flavor combinations are listed in 
Tables \ref {ccnnnn}-\ref {ccnnss}. 
Specifically: 
$R_{12}$ and $R_{34}$ describe the average distance between two charm quarks ($cc$) and two diquarks ($[qq][qq]$), respectively.
$R_{13}$, $R_{14}$, $R_{23}$, and $R_{24}$ represent the average distance between the charm quark $c$ and the diquark $[qq]$.
$R_{12-34}$ stands for the average distance between the mass centers of charm quark pair ($cc$) and the diquark pair ($[qq][qq]$).
$R_{13-24}$ and $R_{14-23}$ represent the average distance between the mass centers of the two pairs of charm quark-diquark $c[qq]$. 
Since the two charm quarks $c$ are identical particles, we have the following relationships:  $R_{13-24}=R_{14-23}$, $R_{13}=R_{23}$, and $R_{14}=R_{24}$. 

If the average distance between the mass centers of the two $c[qq]$ pairs 
 ($R_{13-24}$, $R_{14-23}$) is comparable to or even smaller than the intra-pair quark distances ($R_{12}$,$R_{13}$,$R_{24}$, etc.), 
it implies
strong spatial overlap between the $c[qq]$ clusters. 
Such a configuration supports the interpretation of a compact hexaquark state, as opposed to a loosely bound molecular system.

\subsubsection{Two-body strong decay}\label{sec233}

\begin{figure}[t]
\includegraphics[width=0.85\linewidth]{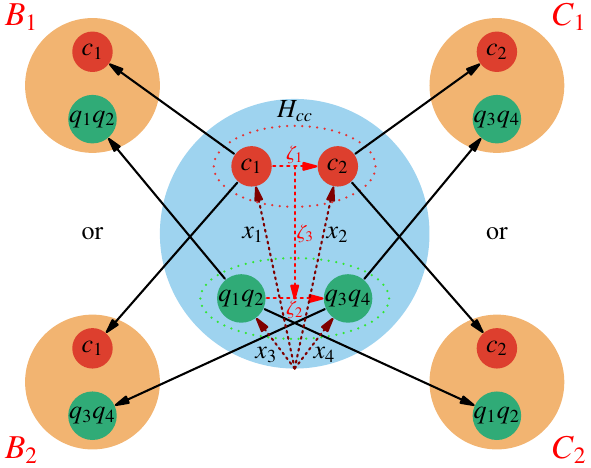}
\caption{
Coordinates defined for the $H_{cc}$ system and its
two-body decays into a baryon pair $BC$ via quark rearrangement. 
Here, the $BC$ final state can form via two quark rearrangement pathways: $B_{1}C_{1}$ ($[c_{1}(q_{1}q_{2})][c_{2}(q_{3}q_{4})]$) and $B_{2}C_{2}$ ($[c_{1}(q_{3}q_{4})][c_{2}(q_{1}q_{2})]$), as illustrated in the figure.
}\label{fig2}
\end{figure}


In addition to mass spectra, we employ the quark-interchange model to calculate the two-body strong decay widths of doubly charmed hexaquarks. 
When the phase space permits, the dominant two-body strong decay for these states is the rearrangement process: $cc[q_{1}q_{2}][q_{3}q_{4}]\to c[q_{1}q_{2}]+c[q_{3}q_{4}]$, as illustrated in Fig. \ref{fig2}. 
Although the decay into a doubly charmed baryon and a light baryon is also kinematically allowed, it is suppressed due to diquark dissociation constraints.
In principle, the light diquarks could both keep well and fall apart in the decay processes. 
However, the two-type assignments of the diquarks have two different decay patterns. 
The symmetry of the diquarks is conserved in the first situation while broken in the second scheme. 
For example, for the excited $\Lambda_c$ states, $D^{(*)}N$ and $\Sigma_c^{(*)}\pi$ are two important decay modes. 
The diquarks of initial $\Lambda_c$ is a whole body in $D^{(*)}N$ channels while broken in $\Sigma_c^{(*)}\pi$ process. 
As shown in Ref.~\cite{Lu:2018utx}, the partial widths of $D^{(*)}N$ are much larger than those of the $\Sigma_c^{(*)}\pi$ channels. 
Similar results are also obtained in Ref.~\cite{Lu:2019rtg}. 
These results imply that in low partial wave, the decays with that diquarks are kept well have a significant tendency. 
Thus, following the experiences in baryon systems, we also consider the decay processes ($cc[q_{1}q_{2}][q_{3}q_{4}]\to c[q_{1}q_{2}]+c[q_{3}q_{4}]$) with that the diquarks are kept well as priority .
Contributions from three-body strong decays and radiative/weak decays are negligible, so we focus exclusively on the above rearrangement channel.

The quark-interchange model \cite{Barnes:1991em,Wong:2001td} describes two-body strong decays via quark rearrangement, driven by the (di)quark-(di)quark interaction $V_{ij}$.
This approach has successfully described decays of exotic states like the $X(3872)$ \cite{Zhou:2019swr}, $X(4630)$ \cite{Yang:2021sue}, $X(2900)$ \cite{Wang:2020prk,Liu:2022hbk}, 
$X(6900)$ \cite{liu:2020eha}, $Z_{c}$ and $Z_{b}$ states \cite{Wang:2018pwi,Xiao:2019spy},
hidden and double charm-strange tetraquark \cite{Liu:2024fnh}, hidden-charm pentaquark $P_{c}$ states \cite{Wang:2019spc},
hidden-charm pentaquarks with triple strangeness \cite{Wang:2021hql}, and all-heavy pentaquark states \cite{Liang:2024met}.

The decay width is given by
\begin{eqnarray}\label{width}
\Gamma=\frac{|\vec{P}_{B}|}{(2J_{A}+1)32\pi^{2}M^{2}_{A}}\int d\Omega|\mathcal{M}(A \to BC)|^{2},
\end{eqnarray}
where $\vec{P}_{B}$ is the final-state three-momentum in the center-of-mass reference frame, and
$M_{A}$ is the initial hexaquark mass.
The decay amplitude $\mathcal{M}(A \to BC)$ 
is \begin{eqnarray}\label{amp}
\mathcal{M}(A \to BC)=
-(2\pi)^{3/2}\sqrt{2M_{A}}\sqrt{2E_{B}}\sqrt{2E_{C}}\times T,
\end{eqnarray}
with  $T$-matrix:
\begin{eqnarray}\label{Eq:T1}
T&=&\sum_{i}\langle\Psi^{B}\Psi^{C}|V_{i}|\Psi^{A}\rangle\nonumber\\
&=&\sum_{i}\langle\Psi^{B}\Psi^{C}|V_{i}|\Psi^{A}_{(cc)}\Psi^{A}_{([qq][qq])}\Psi^{A}_{(cc)-([qq][qq])}\rangle.\quad
\end{eqnarray}
Here, $\Psi^{A}$, $\Psi^{B}$, $\Psi^{C}$ denote the spatial wave functions of the initial hexaquark and final baryons.

\begin{figure*}[htbp]
\begin{tabular*}{0.9\textwidth}{@{\extracolsep{\fill}}cc}
\includegraphics[width=0.4\textwidth]{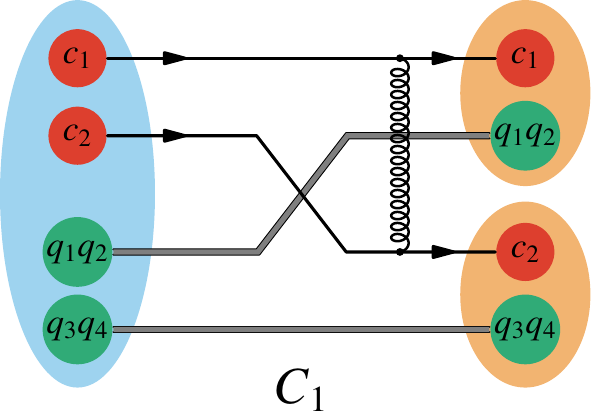}&\includegraphics[width=0.4\textwidth]{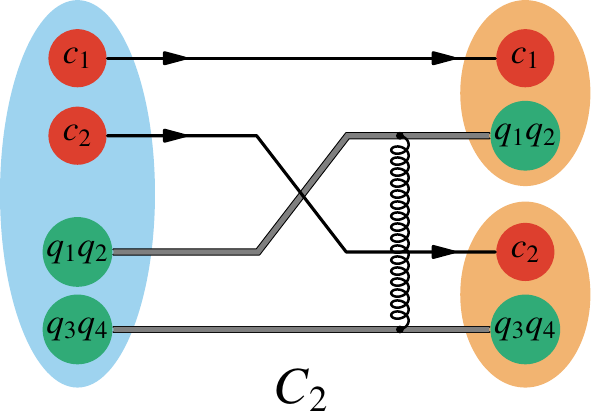}\\
\addlinespace[2em]
\includegraphics[width=0.4\textwidth]{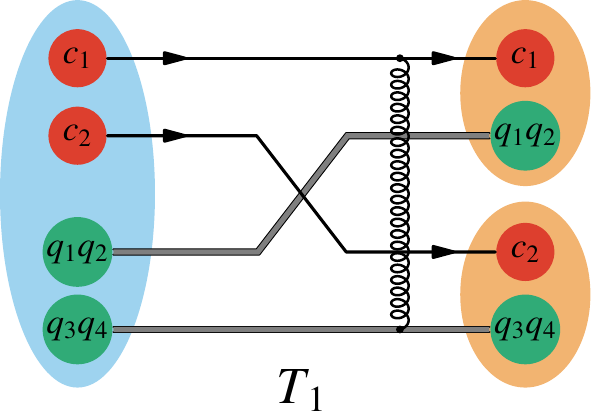}&\includegraphics[width=0.4\textwidth]{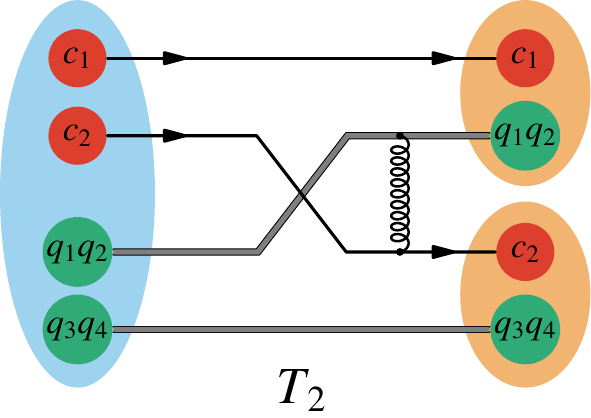}\\
\end{tabular*}
\caption{
The quark-interchange diagrams for $H_{cc}$ decaying into two baryons at the quark level.
The curve line denotes the (di)quark–(di)quark interactions.
}\label{fig3}
\end{figure*}

The $T$-matrix in momentum space integrates the effective potential:
\begin{eqnarray}\label{Eq:T2}
T=\frac{1}{(2\pi)^{3}}\int d\vec{P}_{\alpha}V_{{\rm eff}}(\vec{P}_{\alpha},\vec{P}_{B})\Psi^{A}_{(cc)-([qq][qq])}(\vec{P}_{\alpha}),
\end{eqnarray}
where $V_{{\rm eff}}(\vec{P}_{\alpha},\vec{P}_{B})$ combines contributions from diagrams $C_{1}$, $C_{2}$, $T_{1}$, and $T_{2}$ in Fig.~\ref{fig3}. It factorizes as
\begin{eqnarray}
V_{{\rm eff}}(\vec{P}_{\alpha},\vec{P}_{B})
=I_{{\rm flavor}}I_{{\rm color}}I_{{\rm spin-space}}.
\end{eqnarray}
Here, the flavor factor $I_{{\rm flavor}}$ is taken as 1 for all diagrams.
The color factor $I_{{\rm color}}$ is:
\begin{eqnarray}\label{eq:icolor}
I_{{\rm color}}=\langle\Psi_{B}\Psi_{C}|\frac{\lambda^{c}_{i}}{2}.\frac{\lambda^{c}_{j}}{2}|\Psi^{\bar{3}_{c}(6_{c})}_{A}(cc)\Psi^{3_{c}(\bar{6}_{c})}_{A}([qq][qq])\rangle.\quad
\end{eqnarray}
Numerical values for each diagram are listed in Table \ref{icolor}.
As for the spin-space factor $I_{{\rm spin-space}}$, it is decoupled for doubly charmed hexaquark ground state.
The spin factor $I_{{\rm spin}}$ is
\begin{eqnarray}\label{ispin}
I_{{\rm spin}}=\langle[\chi^{B}_{s_{3}}\chi^{C}_{s_{4}}]_{S'} |\hat{\mathcal{O}}_{s}|[\chi(cc)^{A}_{s_{1}}\chi([qq][qq])^{A}_{s_{2}}]_{S}\rangle,
\end{eqnarray}
where $s_{1}/s_{2}$ and $s_{3}/s_{4}$ represent the spins of initial and final components, and $S$ and $S'$ represent the total spin of initial and final state.
$\hat{\mathcal{O}}_{s}$ stands for the spin operator, taking $\textbf{1}$ for the Coulomb and linear confinement potential, and $\frac{\sigma_{i}}{2}\cdot\frac{\sigma_{j}}{2}$ for the hyperfine potential.
The space factor $I_{{\rm space}}$ is:
\begin{eqnarray}\label{ispace}
&&I_{{\rm space}}=\langle\Psi^{B}\Psi^{C}|\hat{\mathcal{O}}_{q}|\Psi^{A}_{(cc)}\Psi^{A}_{([qq][qq])}\rangle \nonumber\\
&&=\int\int d\textbf{k}_{1}d\textbf{k}_{2}\Psi^{B}(k_{B}+K_{B})\Psi^{C}(k_{C}+K_{C})\hat{\mathcal{O}}_{q}(k_{1}-k_{2})\nonumber\\
&&\quad \Psi^{*A}_{(cc)}(k_{\alpha}+K_{\alpha})\Psi^{*A}_{([qq][qq])}(k_{\beta}+K_{\beta}),
\end{eqnarray}
where $\hat{\mathcal{O}}_{q}$ represents the spatial operator. Its corresponding specific forms are as follows:  $1/q^{2}$, $1/q^{4}$, and $\exp[-q^{2}]$ represent the Coulomb, linear confinement, and hyperfine potential, respectively.
The $\vec{k}_{1}$ ($\vec{k}_{2}$) is the initial (final) three-momenta of the scattered constituent.
We use the $\vec{P}_{B}$,  $\vec{P}_{C}$,  $\vec{P}_{\alpha}$, and  $\vec{P}_{\beta}$ to present the  three-momenta of  final singly-charmed baryons $B$, $C$ and initial doubly charmed hexaquark ($cc$) and ($[qq][qq]$) components, respectively.
For simplicity, we deal with the scattering problem in the center-of-mass frame, so that $\vec{P}_{B}=-\vec{P}_{C}$ and $\vec{P}_{\alpha}=-\vec{P}_{\beta}$.
By applying the above relationships,  for the four quark exchange diagrams in Fig.~\ref{fig3}, 
the relationships of $\vec{k}_{i}$ and $\vec{K}_{i}$ ($i=B, C, \alpha, \beta$) expressed in terms of $\vec{P}_{B}$, $\vec{P}_{\alpha}$, $\vec{k}_{1}$, and $\vec{k}_{2}$ are shown in Table \ref{icolor}.
Here, the constituent quark mass-dependent function $f_{i}$ ($i= \alpha, \beta_{1}, \beta_{2}, B, C$) is:
\begin{eqnarray}
f_{\alpha}&=&\frac{m_{c}}{m_{c}+m_{c}}=\frac{1}{2}, \quad f_{\beta_{1}}=\frac{m_{[q_{1}q_{2}]}}{m_{[q_{1}q_{2}]}+m_{[q_{3}q_{4}]}}, \nonumber \\
f_{\beta_{2}}&=&\frac{m_{[q_{3}q_{4}]}}{m_{[q_{1}q_{2}]}+m_{[q_{3}q_{4}]}},\quad 
f_{B}=\frac{m_{[q_{1}q_{2}]}}{m_{c}+m_{[q_{1}q_{2}]}}, \nonumber \\
f_{C}&=&\frac{m_{[q_{3}q_{4}]}}{m_{c}+m_{[q_{3}q_{4}]}}.
\end{eqnarray}
The specific derivation of integral simplification for Eq. (\ref{ispace}) is referred to Refs.~\cite{Barnes:1991em,Wong:2001td}.
Finally, with the calculated $T$-matrix element $T_{fi}$, the two-body decay widths are computed using Eq.~(\ref{width}) and presented in Tables \ref{ccnnnn}-\ref{ccnnss}.

\begin{table*}[t]
\centering \caption{ The color matrix element $I_{\rm color}=\langle\frac{\lambda^{c}_{i}}{2}.\frac{\lambda^{c}_{j}}{2}\rangle$ and the momentum substitutions in $I_{\rm space}$ for different scattering diagrams.
}\label{icolor}
\begin{lrbox}{\tablebox}
\renewcommand\arraystretch{2}
\renewcommand\tabcolsep{3.5 pt}
\begin{tabular}{c!{\vrule width 0.75pt}cc!{\vrule width 0.75pt}c|c!{\vrule width 0.75pt}c|c!{\vrule width 0.75pt}c|c!{\vrule width 0.75pt}c|c}
\toprule[1.50pt]
\toprule[0.50pt]
\multirow{2}*{Diagram}&\multicolumn{2}{c!{\vrule width 0.75pt}}{$\langle\frac{\lambda^{c}_{i}}{2}.\frac{\lambda^{c}_{j}}{2}\rangle$}&\multicolumn{8}{c}{momentum substitutions}\\
\Xcline{2-11}{0.75pt}
&$(cc)^{\bar{3}_{c}}([qq][qq])^{3_{c}}$&$(cc)^{6_{c}}([qq][qq])^{\bar{6}_{c}}$&$\vec{k}_{\alpha}$&$\vec{K}_{\alpha}$&$\vec{k}_{\beta}$&$\vec{K}_{\beta}$&$\vec{k}_{B}$&$\vec{K}_{B}$&$\vec{k}_{C}$&$\vec{K}_{C}$\\
\Xcline{1-11}{0.75pt}
$C_{1}$&$\frac{2\sqrt{3}}{9}$&$\frac{\sqrt{6}}{9}$&$\vec{k}_{1}$&$-f_{\alpha}\vec{P}_{\alpha}$&$\vec{k}_{2}$&$-f_{\beta 2}\vec{P}_{\alpha}-\vec{P}_{B}$
&$\vec{k}_{2}$&$-f_{B}\vec{P}_{B}$&$\vec{k}_{2}$&$-f_{C}\vec{P}_{B}-\vec{P}_{\alpha}$\\
$C_{2}$&$\frac{2\sqrt{3}}{9}$&$\frac{\sqrt{6}}{9}$&$\vec{k}_{1}$&$-f_{\alpha}\vec{P}_{\alpha}$&$\vec{k}_{2}$&$-f_{\beta 1}\vec{P}_{\alpha}+\vec{P}_{B}$
&$\vec{k}_{1}$&$f_{B}\vec{P}_{B}-\vec{P}_{\alpha}$&$\vec{k}_{1}$&$f_{\beta 1}\vec{P}_{\alpha}$\\
$T_{1}$&$-\frac{2\sqrt{3}}{9}$&$-\frac{\sqrt{6}}{9}$&$\vec{k}_{1}$&$-f_{\alpha}\vec{P}_{\alpha}$&$\vec{k}_{2}$&$-f_{\beta 2}\vec{P}_{\alpha}-\vec{P}_{B}$&$\vec{k}_{2}$&$-f_{\alpha}\vec{P}_{\alpha}$&$\vec{k}_{1}$&$-f_{C}\vec{P}_{B}-\vec{P}_{\alpha}$\\
$T_{2}$&$-\frac{2\sqrt{3}}{9}$&$-\frac{\sqrt{6}}{9}$&$\vec{k}_{1}$&$-f_{\alpha}\vec{P}_{\alpha}$&$\vec{k}_{2}$&$-f_{\beta 1}\vec{P}_{\alpha}+\vec{P}_{B}$&
$\vec{k}_{1}$&$f_{B}\vec{P}_{B}-\vec{P}_{\alpha}$&$\vec{k}_{2}$&$f_{\beta 1}\vec{P}_{\alpha}$\\
\toprule[0.50pt]
\toprule[1.50pt]
\end{tabular}
\end{lrbox}\scalebox{0.95}{\usebox{\tablebox}}
\end{table*}

\section{Numerical results and discussions}\label{sec3}

By solving the  Schr\"{o}dinger equation Eq.~(\ref{tven}) with the Gaussian expansion method, we systematically calculated the mass spectra of the doubly charmed hexaquark system.
The mass range of the doubly charmed hexaquark system approximately spans 5000–6000 MeV.
The mass spectra and relevant configurations are listed in Tables \ref{ccnnnn}-\ref{ccnnss}.
The internal contributions from each part of the Hamiltonian: kinetic energy $\langle T \rangle$,  confinement potential $\langle V^{\rm Con} \rangle$, and hyperfine
interaction potential $\langle V^{\rm SS} \rangle$, are also presented in the same tables.
The results show that the kinetic energy $\langle T \rangle$ and confinement potential $\langle V^{\rm Con} \rangle$ are of the same order of magnitude.
Additionally, it is found that the contribution of hyperfine interaction potential $\langle V^{\rm SS} \rangle$—proportional to $(1/m_{i}m_{j})$—significantly suppresses its mass contribution.
Owing to its smaller mass contribution, the degree of configuration mixing remains relatively low. 
As a result, configuration mixing does not induce obvious mass shifts in physical states.
Thus, each original configuration maintains high purity in its corresponding physical state.
Meanwhile, the mass gaps between different color-spin configurations of the same color configuration ($|\bar{3}3\rangle_{c}$ or $|6\bar{6}\rangle_{c}$) are relatively small, leading to the existence of some partner states in the 
doubly charmed hexaquark system.

Besides the mass spectra and internal mass contributions,
we provide the corresponding RMS radii according to Eq. (\ref{rms}).
Based on the results from relevant tables, most of RMS radii are in the range from 1.2 to 1.6 fm, which are roughly the same order of magnitude. 
If it is a molecular-state configuration, some distances between charm quark ($c$) and diquark ($[qq]$) are much greater than those between the two charm quarks ($cc$) and between the two diquarks ($[qq]-[qq]$). 
Meanwhile, the RMS radius of the molecular configuration can reach several femtometers.
Therefore, our calculation results are consistent with the expectations of the compact hexaquark configuration.
Finally, based on Eq. (\ref{width}), we presented the partial widths of each state decaying into different two-body final states, as well as its total width.
It should be noted that the total decay width here ignores the suppressed three-body strong decays, the two-body strong decays with the final states: doubly charmed baryon+light-flavor baryon ($ccq+qqq$), as well as the radiative decays and weak decays. 
Therefore, the actual total widths of these states will be slightly larger than our calculated values.

For clarity, according to Tables \ref{ccnnnn}-\ref{ccnnss}, the relative mass positions of each state, the total decay widths, and the corresponding rearrangement decay channels are plotted in
Figs. \ref{fig-ccnnnn}-\ref{fig-ccnnss}. 
For convenience,  we also label all possible spin (isospin) quantum numbers of the rearrangement decay channels with subscripts (superscripts).
According to the above-mentioned figures, there is no stable state in the doubly charmed hexaquark system,
and they are all unstable states which can decay into two singly-charmed baryons through two-body strong interaction.
The reason is that the pairwise attractive interaction provided by $\langle V^{\rm Con} \rangle$ is far smaller than that of the two singly-charmed baryons in the decay final states. Therefore, their masses are higher than the threshold of the decay final states.
For simplicity, we use the notation ${\rm H_{content}}({\rm Mass}, I, J^{P})$ to label a particular doubly charmed hexaquark state.
\\
\subsection{The $cc[nn][nn]$ and $cc[ss][ss]$ subsystems}

\begin{table*}[t]
\centering
\caption{
The numerical results of the mass spectrum, the mass contributions of each Hamiltonian part (in MeV), the root-mean-square radii (in fm), and the partial decay widths and total decay widths of the fall-apart decay processes (in MeV) for the $cc[nn][nn]$ and $cc[ss][ss]$ hexaquark states. 
}\label{ccnnnn}
\begin{lrbox}{\tablebox}
\renewcommand\arraystretch{1.9}
\renewcommand\tabcolsep{0.75 pt}
\begin{tabular}{ccc|ccc|cccccc|rrrrrr|r}
\toprule[1.50pt]
\toprule[0.50pt]
\multicolumn{3}{l|}{$cc[nn][nn]$}&
\multicolumn{3}{c|}{Internal contribution}& 
\multicolumn{6}{c|}{RMS Radius}&\multicolumn{7}{c}{Fall-apart decay properties}\\
\Xcline{4-19}{0.3pt}
\multirow{2}*{$I[J^{P}]$}&\multirow{2}*{Configuration}&\multirow{2}*{Mass}&\multirow{2}*{$\langle T \rangle$}
&\multirow{2}*{$\langle V^{\rm Con} \rangle$}
&\multirow{2}*{$\langle V^{\rm SS} \rangle$}
&\multirow{2}*{$R_{12}$}&\multirow{2}*{$R_{34}$}
&$R_{13}$&$R_{23}$
&\multirow{2}*{$R_{12-34}$}
&$R_{13-24}$
&\multirow{2}*{$\Sigma^{*}_{c}\Sigma^{*}_{c}$}
&\multirow{2}*{$\Sigma^{*}_{c}\Sigma_{c}$}
&\multirow{2}*{$\Sigma_{c}\Sigma_{c}$}
&\multirow{2}*{$\Lambda_{c}\Sigma^{*}_{c}$}
&\multirow{2}*{$\Lambda_{c}\Sigma_{c}$}
&\multirow{2}*{$\Lambda_{c}\Lambda_{c}$}
&\multicolumn{1}{c}{\multirow{2}*{$\Gamma_{sum}$}}\\
&&&&&&&&$R_{14}$&$R_{24}$&&$R_{14-23}$&&&&&&&\\
\bottomrule[1.00pt]
\multirow{2}*{$2(0)[2^{+}]$}&
\multirow{2}*{$\begin{pmatrix}
|[(nn)^{I=1,\bar{3}_{c}}_{s=1}(nn)^{I=1,\bar{3}_{c}}_{s=1}]^{I=2,3_{c}}_{s=1}(cc)^{I=0,\bar{3}_{c}}_{s=1}\rangle^{I=2}_{s=2}
\\
|[(nn)^{I=1,\bar{3}_{c}}_{s=1}(nn)^{I=1,\bar{3}_{c}}_{s=1}]^{I=2,\bar{6}_{c}}_{s=2}(cc)^{I=0,6_{c}}_{s=0}\rangle^{I=2}_{s=2}
\end{pmatrix}$}
&
\multirow{2}*{$\begin{pmatrix}5598\\5504\end{pmatrix}$}&1377.9&-1571.9&3.9&1.29&1.26&1.36&1.36&1.44&1.28&29.0&0.1&&&&&29.0
\\
&&&1369.7&-1656.9&3.6&1.49&1.47&1.31&1.31&1.12&1.48&19.3&19.4&&&&&38.7
\\
\toprule[0.05pt]
\multirow{1}*{$2(0)[1^{+}]$}&$|[(nn)^{I=1,\bar{3}_{c}}_{s=1}(nn)^{I=1,\bar{3}_{c}}_{s=1}]^{I=2,3_{c}}_{s=1}(cc)^{I=0,\bar{3}_{c}}_{s=1}\rangle^{I=2}_{s=1}$&
5577&1389.3&-1583.2&-17.1&1.29&1.25&1.35&1.35&1.43&1.27&6.0&0.7&28.1&&&&34.8
\\
\toprule[0.05pt]
\multirow{3}*{$2(0)[0^{+}]$}&
\multirow{3}*{$\begin{pmatrix}
|[(nn)^{I=1,\bar{3}_{c}}_{s=1}(nn)^{I=1,\bar{3}_{c}}_{s=1}]^{I=2,3_{c}}_{s=1}(cc)^{I=0,\bar{3}_{c}}_{s=1}\rangle^{I=2}_{s=0}
\\
|[(nn)^{I=1,\bar{3}_{c}}_{s=1}(nn)^{I=1,\bar{3}_{c}}_{s=1}]^{I=2,\bar{6}_{c}}_{s=0}(cc)^{I=0,6_{c}}_{s=0}\rangle^{I=2}_{s=0}
\\
|[(nn)^{I=0,\bar{3}_{c}}_{s=0}(nn)^{I=0,\bar{3}_{c}}_{s=0}]^{I=0,\bar{6}_{c}}_{s=0}(cc)^{I=0,6_{c}}_{s=0}\rangle^{I=0}_{s=0}
\end{pmatrix}$}
&
\multirow{3}*{$\begin{pmatrix}5582\\5512\\5043\end{pmatrix}$}
&1386.4&-1580.4&-11.8&1.29&1.26&1.35&1.35&1.43&1.27&23.3&&21.0&&&&44.3\\
&&&1365.3&-1652.5&11.3&1.49&1.47&1.31&1.31&1.13&1.48&22.9&&4.1&&&&27.0
\\
\cdashline{1-19}[1pt/1pt]
$0[0^{+}]$&&&1389.5&-1471.5&5.8&1.63&1.52&1.40&1.40&1.20&1.58&&&&&&14.7&14.7
\\
\bottomrule[0.90pt]
\multirow{1}*{$1[3^{+}]$}&$|[(nn)^{I=1,\bar{3}_{c}}_{s=1}(nn)^{I=1,\bar{3}_{c}}_{s=1}]^{I=1,3_{c}}_{s=2}(cc)^{I=0,\bar{3}_{c}}_{s=1}\rangle^{I=1}_{s=3}$
&5590&1372.3&1632.8&-4.1&1.30&1.26&1.36&1.36&1.43&1.27&36.7&&&&&&36.7
\\
\toprule[0.05pt]
\multirow{2}*{$1[2^{+}]$}&
\multirow{2}*{$\begin{pmatrix}
|[(nn)^{I=1,\bar{3}_{c}}_{s=1}(nn)^{I=1,\bar{3}_{c}}_{s=1}]^{I=1,3_{c}}_{s=2}(cc)^{I=0,\bar{3}_{c}}_{s=1}\rangle^{I=1}_{s=2}\\
|[(nn)^{I=0,\bar{3}_{c}}_{s=0}(nn)^{I=1,\bar{3}_{c}}_{s=1}]^{I=1,3_{c}}_{s=1}(cc)^{I=0,\bar{3}_{c}}_{s=1}\rangle^{I=1}_{s=2}
\end{pmatrix}$}
&
\multirow{2}*{$\begin{pmatrix}5600\\5375\end{pmatrix}$}
&1376.6&-1570.6&6.4&1.30&1.26&1.36&1.36&1.43&1.28&4.8&9.5&&&&&15.3
\\
&&&1383.5&-1475.0&10.4&1.37&1.27&1.45&1.36&1.27&1.48&&&&70.1&&&70.1
\\
\toprule[0.05pt]
\multirow{5}*{$1[1^{+}]$}&
\multirow{5}*{$\begin{pmatrix}|[(nn)^{I=1,\bar{3}_{c}}_{s=1}(nn)^{I=1,\bar{3}_{c}}_{s=1}]^{I=1,\bar{6}_{c}}_{s=1}(cc)^{I=0,6_{c}}_{s=0}\rangle^{I=1}_{s=1}
\\
|[(nn)^{I=1,\bar{3}_{c}}_{s=1}(nn)^{I=1,\bar{3}_{c}}_{s=1}]^{I=1,3_{c}}_{s=2}(cc)^{I=0,\bar{3}_{c}}_{s=1}\rangle^{I=1}_{s=1}
\\
|[(nn)^{I=1,\bar{3}_{c}}_{s=1}(nn)^{I=1,\bar{3}_{c}}_{s=1}]^{I=1,3_{c}}_{s=0}(cc)^{I=0,\bar{3}_{c}}_{s=1}\rangle^{I=1}_{s=1}
\\
|[(nn)^{I=0,\bar{3}_{c}}_{s=0}(nn)^{I=1,\bar{3}_{c}}_{s=1}]^{I=1,3_{c}}_{s=1}(cc)^{I=0,\bar{3}_{c}}_{s=1}\rangle^{I=1}_{s=1}
\\
|[(nn)^{I=0,\bar{3}_{c}}_{s=0}(nn)^{I=1,\bar{3}_{c}}_{s=1}]^{I=1,\bar{6}_{c}}_{s=1}(cc)^{I=0,6_{c}}_{s=0}\rangle^{I=1}_{s=1}
\end{pmatrix}$}
&
\multirow{5}*{$\begin{pmatrix}5590\\5586\\5510\\5365\\5283\end{pmatrix}$}
&1382.3&-1576.3&-4.1&1.30&1.26&1.36&1.36&1.43&1.27&5.7&0.6&26.6&&&&32.9
\\
&&&1384.3&-1578.2&-7.8&1.29&1.26&1.36&1.36&1.44&1.27&19.6&0.4&0.8&&&&20.8
\\
&&&1366.8&-1654.0&8.7&1.49&1.47&1.31&1.31&1.13&1.48&19.4&2.2&2.8&&&&24.4
\\
&&&1389.0&-1480.5&-0.1&1.37&1.26&1.45&1.36&1.48&1.31&&&&7.8&3.5&&11.3
\\
&&&1376.3&-1558.8&9.6&1.56&1.49&1.42&1.30&1.16&1.52&&&&15.4&2.7&&18.1
\\
\bottomrule[1.50pt]
\multicolumn{3}{l|}{$cc[ss][ss]$}&
\multicolumn{3}{c|}{Internal contribution}& 
\multicolumn{6}{c|}{RMS Radius}&\multicolumn{7}{c}{Fall-apart decay properties}\\
\Xcline{4-19}{0.3pt}
\multirow{2}*{$I[J^{P}]$}&\multirow{2}*{Configuration}&\multirow{2}*{Mass}&\multirow{2}*{$\langle T \rangle$}
&\multirow{2}*{$\langle V^{\rm Con} \rangle$}
&\multirow{2}*{$\langle V^{\rm SS} \rangle$}
&\multirow{2}*{$R_{12}$}&\multirow{2}*{$R_{34}$}
&$R_{13}$&$R_{23}$
&\multirow{2}*{$R_{12-34}$}
&$R_{13-24}$
&\multicolumn{2}{c}{\multirow{2}*{$\Omega^{*}_{c}\Omega^{*}_{c}$}}
&\multicolumn{2}{c}{\multirow{2}*{$\Omega^{*}_{c}\Omega_{c}$}}
&\multicolumn{2}{c|}{\multirow{2}*{$\Omega_{c}\Omega_{c}$}}
&\multicolumn{1}{c}{\multirow{2}*{$\Gamma_{sum}$}}\\
&&&&&&&&$R_{14}$&$R_{24}$&&$R_{14-23}$&&&&&&&\\
\bottomrule[1.00pt]
\multirow{2}*{$0[2^{+}]$}&
\multirow{2}*{$\begin{pmatrix}|[(ss)^{I=0,\bar{3}_{c}}_{s=1}(ss)^{I=0,\bar{3}_{c}}_{s=1}]^{I=0,3_{c}}_{s=1}(cc)^{I=0,\bar{3}_{c}}_{s=1}\rangle^{I=0}_{s=2}\\
|[(ss)^{I=0,\bar{3}_{c}}_{s=1}(ss)^{I=0,\bar{3}_{c}}_{s=1}]^{I=0,\bar{6}_{c}}_{s=2}(cc)^{I=0,6_{c}}_{s=0}\rangle^{I=0}_{s=2}
\end{pmatrix}$}&
\multirow{2}*{$\begin{pmatrix}6038\\5945\end{pmatrix}$}
&1370.4&-1692.1&5.4&1.20&1.25&1.30&1.30&1.38&1.22&\multicolumn{2}{c}{40.0}&\multicolumn{2}{c}{1.0}&\multicolumn{2}{c|}{}&41.0
\\
&&&1361.9&-1775.8&4.8&1.40&1.43&1.26&1.26&1.08&1.42&\multicolumn{2}{c}{15.8}&\multicolumn{2}{c}{10.4}&\multicolumn{2}{c|}{}&26.2
\\
\toprule[0.05pt]
\multirow{1}*{$0[1^{+}]$}&$|[(ss)^{I=0,\bar{3}_{c}}_{s=1}(ss)^{I=0,\bar{3}_{c}}_{s=1}]^{I=0,3_{c}}_{s=1}(cc)^{I=0,\bar{3}_{c}}_{s=1}\rangle^{I=0}_{s=1}$&
6021&1379.5&-1701.2&-11.3&1.19&1.24&1.30&1.30&1.37&1.22&\multicolumn{2}{c}{2.1}&\multicolumn{2}{c}{9.1}&\multicolumn{2}{c|}{0.7}&11.9
\\
\toprule[0.05pt]
\multirow{2}*{$0[0^{+}]$}&
\multirow{2}*{$\begin{pmatrix}|[(ss)^{I=0,\bar{3}_{c}}_{s=1}(ss)^{I=0,\bar{3}_{c}}_{s=1}]^{I=0,3_{c}}_{s=1}(cc)^{I=0,\bar{3}_{c}}_{s=1}\rangle^{I=0}_{s=0}
\\
|[(ss)^{I=0,\bar{3}_{c}}_{s=1}(ss)^{I=0,\bar{3}_{c}}_{s=1}]^{I=0,\bar{6}_{c}}_{s=0}(cc)^{I=0,6_{c}}_{s=0}\rangle^{I=0}_{s=0}
\end{pmatrix}$}
&
\multirow{2}*{$\begin{pmatrix}6025\\5950\end{pmatrix}$}
&1377.3&-1698.9&-7.1&1.19&1.25&1.30&1.30&1.37&1.22&\multicolumn{2}{c}{27.7}&\multicolumn{2}{c}{}&\multicolumn{2}{c|}{35.5}&63.2
\\
&&&1359.3&-1773.2&9.5&1.40&1.43&1.26&1.26&1.08&1.42&\multicolumn{2}{c}{31.0}&\multicolumn{2}{c}{}&\multicolumn{2}{c|}{8.7}&39.7
\\
\bottomrule[0.50pt]
\bottomrule[1.50pt]
\end{tabular}
\end{lrbox}\scalebox{0.78}{\usebox{\tablebox}}
\end{table*}

\begin{figure*}[htbp]
\begin{tabular}{c}
\includegraphics[width=\textwidth]{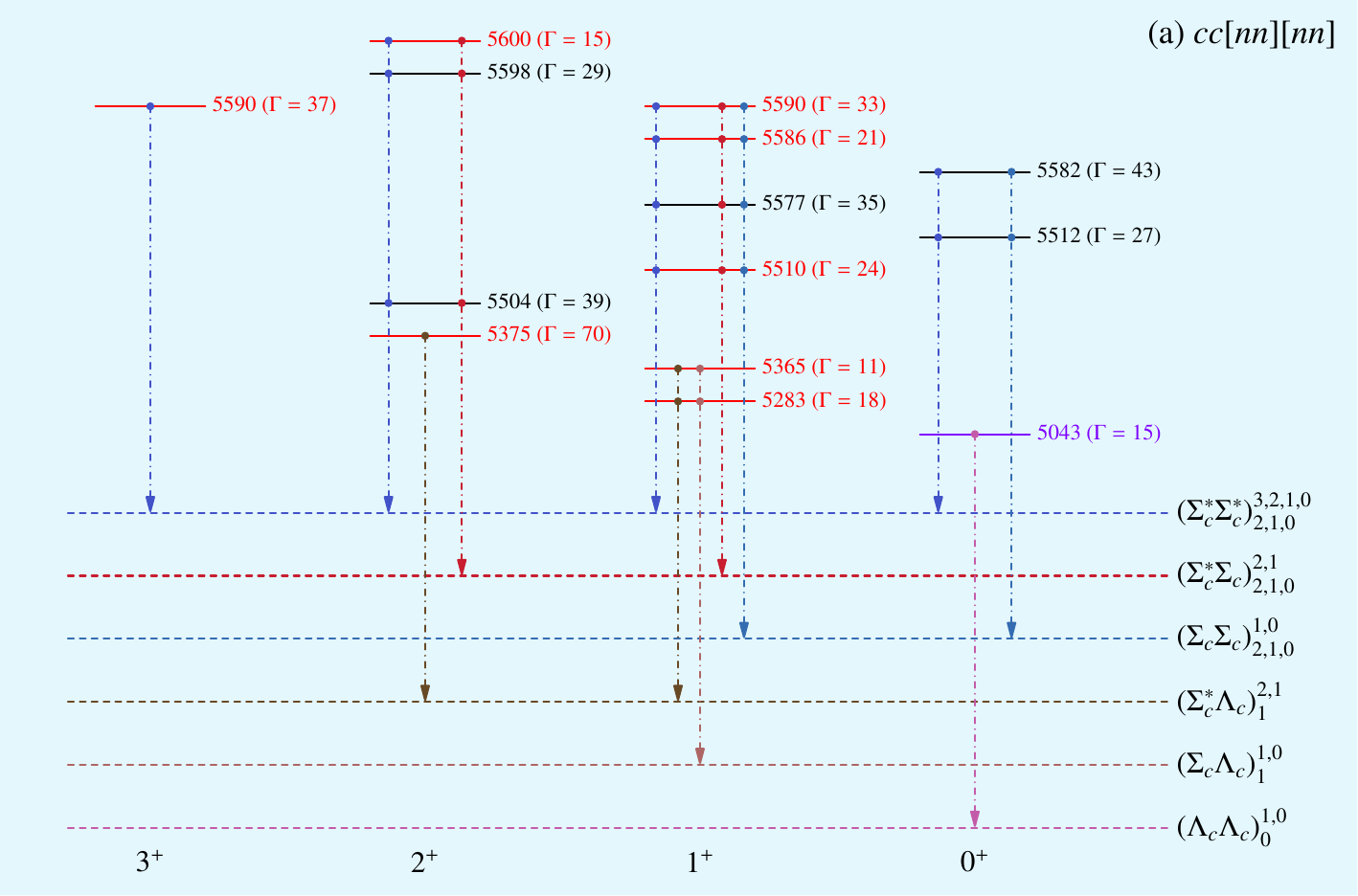}\\
\includegraphics[width=\textwidth]{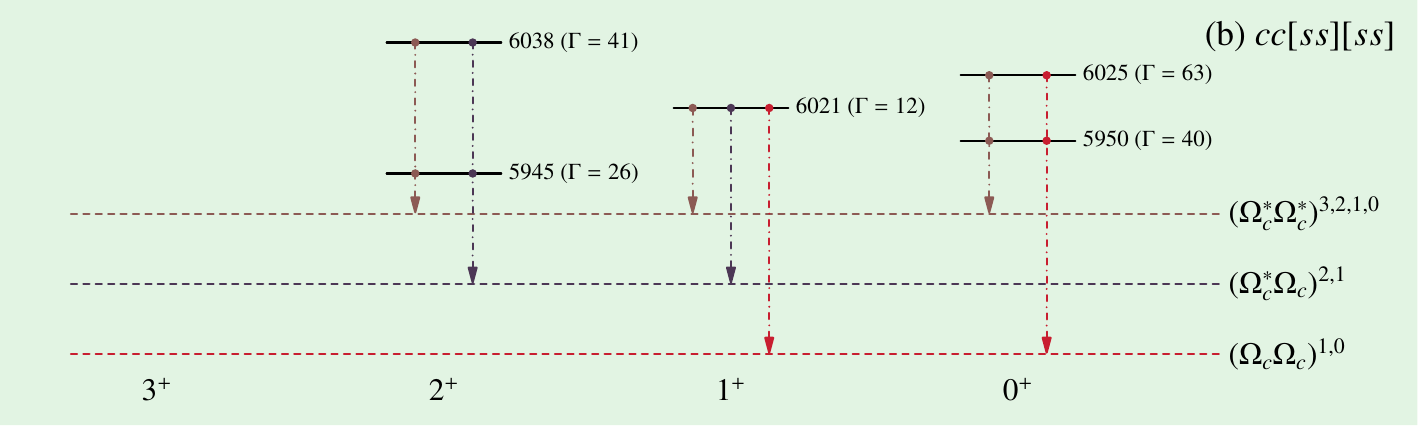}\\
\end{tabular}
\caption{
Relative positions for the $cc[nn][nn]$ (a) and $cc[ss][ss]$ (b) hexaquark states labeled with horizontal solid lines, the labels, e.g. $5582~(\Gamma=43)$ represents the mass and total decay width of the corresponding state (units: MeV) .
In the $cc[nn][nn]$ subsystem, the black, red, and purple horizontal lines represent the hexaquark states with $I=2,0$, $I=1$, and $I=0$, respectively.
The dotted lines denote various S-wave baryon-baryon thresholds, and the superscripts (subscript) of the labels, e.g. $(\Sigma^{*}_{c}\Sigma^{*}_{c})^{3,2,1,0}_{2,1,0}$, represent the possible total angular momenta (isospin) of the channels.
The solid dots of different colors where the vertical dashed lines with arrows intersect the horizontal solid lines represent the allowed rearranged S-wave decay processes. 
If a vertical dashed line with an arrow intersects a horizontal solid line without a solid dot, it means that corresponding decay process is forbidden for relevant state.
}\label{fig-ccnnnn}
\end{figure*}

In the following, we first discuss the $cc[nn][nn]$ and $cc[ss][ss]$ subsystems.
For the $cc[nn][nn]$ subsystem, the isospin of two $[nn]$ diquarks can couple to $I=2, 1,$ and $0$.
According to the Spin-Statistics Theorem, the $I=2$ states and the $I=0$ states have the same symmetry constraints for total wave functions, leading to their degeneracy in mass spectra, RMS radii, and decay properties.
A similar situation also occurs in doubly-heavy pentaquark system.
For the $ccnn\bar{n}$ and $bbnn\bar{n}$ subsystems with $I_{nn} = 1$, the same spectra are obtained for the case of the total isospin 
$I=1/2$ and $3/2$ \cite{Zhou:2018bkn}.
This arises because both the confinement potential and hyperfine interaction in the Hamiltonian are independent of isospin.

Of course, there is also a special configuration:  
$|[(nn)^{I=0}_{s=0}(nn)^{I=0}_{s=0}]^{I=0,\bar{6}_{c}}_{s=0}(cc)^{I=0,6_{c}}_{s=0}\rangle^{I=0}_{s=0}$.
In this configuration,  the two $[nn]$ diquarks only couple to total isospin  $I=0$.
This feature leads to an additional allowed state:  ${\rm H}_{c^{2}[nn]^{2}}(5043, 0, 0^+)$ in the 
 $I(J^P)=0(0^+)$  state,  which is absent in the $I(J^P)=0(2^+)$  sector. 
Notably, due to the significant suppression of color-spin mixing, the mass gaps between $I(J^P)=0(0^+)$ and $0(2^+)$ states  are negligible ($\Delta m<1$ MeV ), maintaining approximate degeneracy. 
Apart from  ${\rm H}_{c^{2}[nn]^{2}}(5043, 0, 0^+)$  state, the remaining $0(0^{+})$ and $2(0^{+})$ states still have the same mass, RMS radii, and decay behavior, and are represented by the same data set in Table \ref{ccnnnn}.

For the ${\rm H}_{c^{2}[nn]^{2}}(5043, 0, 0^{+})$ state,  because of the low mass and narrow width, it may be the most ideal in the doubly charmed hexaquark system. In this scheme, it is potential to observe this state. 
Its unique properties are as follows:
first, both \([nn]\) diquarks in the state are scalar diquarks with $I=0$, $S=0$. 
Compared with other diquark configurations, the structures of this system have an antisymmetric color part and an antisymmetric spin part, which result in the strongest internal confinement potential and chromomagnetic interaction, and lead to higher stability. 
Consequently, this configuration achieves the lowest mass (5043 MeV) within the doubly charmed hexaquark spectrum.  
Second, this state mainly decays into the $\Lambda_c\Lambda_c$ final state, while other decay channels are significantly suppressed. Since the $\Lambda_c$ is well established, we strongly suggest searching for hexaquark state in the $\Lambda_c\Lambda_c$ channel.
Finally, its total width $\Gamma_{\rm total} = 14.3$ MeV.
Though it is wider than the sub-MeV-scale width of the experimentally discovered \(T_{cc}^+(3875)\), it still remains characteristic of a narrow hadronic resonance.
This is because \(T_{cc}^+(3875)\) lies slightly below the $D^{*+}D^0$ threshold, making the three-body decay $D^0D^0\pi$ its dominant channel and resulting in an extremely narrow width. In contrast, ${\rm H}_{c^{2}[nn]^{2}}(5043, 0, 0^+)$ state is above the $\Lambda_c\Lambda_c$ threshold, with a larger decay phase space.

The discovery of \(T_{cc}^+(3875)\) again demonstrates the experimental detectability of the doubly charmed multi-quark system. Therefore, we suggest that experimental collaborations like LHCb, CMS, ATLAS, etc. produce the $\Lambda_c\Lambda_c$ final state via high-luminosity $pp$ collisions and analyze the invariant mass spectrum within the $5000$–$5100$ MeV range to search for this narrow peak structure.

For other two $I[J^{P}]=2(0)[0^{+}]$ states:  ${\rm H}_{c^{2}[nn]^{2}}(5582,\\ 2(0), 0^{+})$ and  ${\rm H}_{c^{2}[nn]^{2}}(5512, 2(0), 0^{+})$, due to large phase space, they both decay into the $\Sigma^{*}_{c}\Sigma^{*}_{c}$ and $\Sigma_{c}\Sigma_{c}$ final states.
Their mass gap reaches 70 MeV, because of different color configurations.
Further, we find that compared with the $\bar{3}\otimes3$ color configuration, the $6\otimes\bar{6}$ color configuration exhibits stronger attractive interactions, resulting in lower masses for these states.
Meanwhile, their total decay widths are 44 MeV and 27 MeV, respectively.
The corresponding relative partial decay width ratios are as follows:
\begin{eqnarray}
\frac{\Gamma[{\rm H}_{c^{2}[nn]^{2}}(5582, 2(0), 0^{+})\to \Sigma_{c}\Sigma_{c}]}{\Gamma[{\rm H}_{c^{2}[nn]^{2}}(5582, 2(0), 0^{+})\to\Sigma^{*}_{c}\Sigma^{*}_{c}]}=1,
\end{eqnarray}
and 
\begin{eqnarray}
\frac{\Gamma[{\rm H}_{c^{2}[nn]^{2}}(5512, 2(0), 0^{+})\to \Sigma_{c}\Sigma_{c}]}{\Gamma[{\rm H}_{c^{2}[nn]^{2}}(5512, 2(0), 0^{+})\to\Sigma^{*}_{c}\Sigma^{*}_{c}]}=\frac{2}{11},
\end{eqnarray}
respectively.
For other $I=2(0)$ $cc[nn][nn]$ states, one can perform similar discussions on the decay behaviors according to Table \ref{ccnnnn} and Fig. \ref{fig-ccnnnn}.

For the $I=1$ $cc[nn][nn]$ states, there is no ground $I[J^{P}]=1[0^{+}]$ state due to symmetry constraints.
Meanwhile, the total decay widths of most of states are larger than 20 MeV, classifying them as relatively broad states.
Among these states, the ${\rm H}_{c^{2}[nn]^{2}}(5365, 1, 1^{+})$ is the narrowest state. Although its decay phase space is larger than that of ${\rm H}_{c^{2}[nn]^{2}}(5283, 1, 1^{+})$, its total width is only about 10 MeV. 
For this state, we have 
\begin{eqnarray}
\frac{\Gamma[{\rm H}_{c^{2}[nn]^{2}}(5365, 1, 1^{+})\to \Sigma_{c}\Lambda_{c}]}{\Gamma[{\rm H}_{c^{2}[nn]^{2}}(5365, 1, 1^{+})\to\Sigma^{*}_{c}\Lambda_{c}]}=\frac{1}{2}.
\end{eqnarray}
This indicates that the $\Sigma^{*}_{c}\Lambda_{c}$ decay channel is dominant.
The narrow width of ${\rm H}_{c^{2}[nn]^{2}}(5365, 1, 1^{+})$ makes its peak shape significantly distinguishable from the background.
Therefore, we suggest that experiments scan the invariant mass spectrum of $\Sigma^{*}_{c}\Lambda_{c}$ in the mass range of $5300$-$5400$ MeV, with particular attention to the narrow peak structure near 5360 MeV. If the peak position and width observed in the experiment are consistent with theoretical predictions, and the branching ratio conforms to the 1:2 ratio, it can be confirmed as the signal of ${\rm H}_{c^{2}[nn]^{2}}(5365, 1, 1^{+})$.

Moreover, for the ${\rm H}_{c^{2}[nn]^{2}}(5590, 1, 1^{+})$ and ${\rm H}_{c^{2}[nn]^{2}}\\(5586, 1, 1^{+})$ states, they are degenerate states, which have same quantum numbers and similar masses ($\Delta M = 4$ MeV).
Although they have similar decay phase space, the total decay width of ${\rm H}_{c^{2}[nn]^{2}}(5590, 1, 1^{+})$ is 33 MeV, approximately 1.5 times that of ${\rm H}_{c^{2}[nn]^{2}}(5586, 1, 1^{+})$.
And their relative partial decay width ratios are
\begin{eqnarray}
\Gamma_{\Sigma^{*}_{c}\Sigma^{*}_{c}}:\Gamma_{\Sigma_{c}\Sigma^{*}_{c}}:\Gamma_{\Sigma_{c}\Sigma_{c}}=1:0.1:5,
\end{eqnarray}
and
\begin{eqnarray}
\Gamma_{\Sigma^{*}_{c}\Sigma^{*}_{c}}:\Gamma_{\Sigma_{c}\Sigma^{*}_{c}}:\Gamma_{\Sigma_{c}\Sigma_{c}}=25:0.5:1,
\end{eqnarray}
respectively.
Evidently, $\Sigma_{c}\Sigma_{c}$ and $\Sigma^{*}_{c}\Sigma^{*}_{c}$ are their dominant decay channels, respectively.
Although theoretically we can distinguish them by their total decay widths and the branching ratios, current experimental detectors still face great difficulties in distinguishing degenerate states with a mass difference ($\Delta M$) of 4 MeV.

For the $cc[ss][ss]$ subsystem, since it has exactly the same symmetry constraints as the $cc[nn][nn]$ subsystem with $I=2$, the number of allowed states is also identical.
Due to symmetry constraints, there is no ground $I[J^{P}]=0[3^{+}]$ state.
Among them, the ${\rm H}_{c^{2}[ss]^{2}}(6021, 0, 1^{+})$ is the narrowest state whose total width is around 12 MeV, even though it has more allowed decay channels: $\Omega_{c}\Omega_{c}$, $\Omega^{*}_{c}\Omega_{c}$, and $\Omega^{*}_{c}\Omega^{*}_{c}$.
Here, we obtain the following relative ratio of decay widths:
\begin{eqnarray}
\Gamma_{\Omega^{*}_{c}\Omega^{*}_{c}}:\Gamma_{\Omega^{*}_{c}\Omega_{c}}:\Gamma_{\Omega_{c}\Omega_{c}}=3:13:1.
\end{eqnarray}
Our results show that the $\Omega^{*}_{c}\Omega_{c}$ channel is its dominant decay channel.
Moreover, all the $cc[ss][ss]$ states can decay into $\Omega^{*}_{c}\Omega^{*}_{c}$ final states, and this decay channel is crucial for identifying $cc[ss][ss]$ states.
Although theoretical predictions indicate clear signals in the $cc[ss][ss]$ subsystem, experimental discovery of this subsystem still faces significant challenges. Firstly, the production probability of the $[ss]$ quark pair in proton-proton collisions is much lower than that of the $[nn]$ pair. Additionally, the experimental reconstruction of its decay final states $\Omega^{*}_{c}$ and $\Omega_{c}$ is relatively complex. 
\\

\subsection{The $cc[nn][ns]$ and $cc[ss][sn]$ subsystems}

\begin{table*}[t]
\centering \caption{
The numerical results of the mass spectrum,
the mass contributions of each Hamiltonian part (in MeV), the root-mean-square radii (in fm), and the partial decay widths and total decay widths of the fall-apart decay processes (in MeV) for the $cc[nn][ns]$ and $cc[ss][sn]$ hexaquark states. 
}\label{ccnnns}
\begin{lrbox}{\tablebox}
\renewcommand\arraystretch{1.75}
\renewcommand\tabcolsep{0.65 pt}
\begin{tabular}{ccc|ccc|cccccc|rrrrrr|r}
\toprule[1.50pt]
\toprule[0.50pt]
\multicolumn{3}{l|}{$cc[nn][ns]$}&
\multicolumn{3}{c|}{Internal contribution}& \multicolumn{6}{c|}{RMS Radius}&\multicolumn{7}{c}{Fall-apart decay properties}\\
\Xcline{4-19}{0.3pt}
\multirow{2}*{$I[J^{P}]$}&\multirow{2}*{Configuration}&\multirow{2}*{Mass}&\multirow{2}*{$\langle T \rangle$}&\multirow{2}*{$\langle V^{\rm Con} \rangle$}&\multirow{2}*{$\langle V^{\rm SS} \rangle$}&\multirow{2}*{$R_{12}$}&\multirow{2}*{$R_{34}$}&$R_{13}$&$R_{23}$
&\multirow{2}*{$R_{12-34}$}
&$R_{13-24}$
&\multirow{2}*{$\Sigma^{*}_{c}\Xi^{*}_{c}$}
&\multirow{2}*{$\Sigma^{*}_{c}\Xi'_{c}$}
&\multirow{2}*{$\Sigma_{c}\Xi^{*}_{c}$}
&\multirow{2}*{$\Sigma_{c}\Xi'_{c}$}
&\multirow{2}*{$\Sigma^{*}_{c}\Xi_{c}$}
&\multirow{2}*{$\Sigma_{c}\Xi_{c}$}
&\multicolumn{1}{c}{\multirow{2}*{$\Gamma_{sum}$}}\\
 &&&&&&&&$R_{14}$&$R_{24}$&&$R_{14-23}$&&&&&&\\
\bottomrule[1.00pt]
\multirow{1}*{$\frac{3}{2}(\frac{1}{2})[3^{+}]$}&
$|[(nn)^{I=1,\bar{3}_{c}}_{s=1}(ns)^{I=\frac{1}{2},\bar{3}_{c}}_{s=1}]^{I=\frac{3}{2},3_{c}}_{s=2}(cc)^{I=0,\bar{3}_{c}}_{s=1}\rangle^{I=\frac{3}{2}}_{s=3}$&5734&1366.6&-1597.1&20.6&1.27&1.26&1.36&1.33&1.42&1.26&42.1&&&&&&42.1
\\
\toprule[0.05pt]
\multirow{4}*{$\frac{3}{2}(\frac{1}{2})[2^{+}]$}&
\multirow{4}*{$\begin{pmatrix}
|[(nn)^{I=1,\bar{3}_{c}}_{s=1}(ns)^{I=\frac{1}{2},\bar{3}_{c}}_{s=1}]^{I=\frac{3}{2},3_{c}}_{s=2}(cc)^{I=0,\bar{3}_{c}}_{s=1}\rangle^{I=\frac{3}{2}}_{s=2}\\
|[(nn)^{I=1,\bar{3}_{c}}_{s=1}(ns)^{I=\frac{1}{2},\bar{3}_{c}}_{s=1}]^{I=\frac{3}{2},3_{c}}_{s=1}(cc)^{I=0,\bar{3}_{c}}_{s=1}\rangle^{I=\frac{3}{2}}_{s=2}\\
|[(nn)^{I=1,\bar{3}_{c}}_{s=1}(ns)^{I=\frac{1}{2},\bar{3}_{c}}_{s=1}]^{I=\frac{3}{2},\bar{6}_{c}}_{s=2}(cc)^{I=0,6_{c}}_{s=0}\rangle^{I=\frac{3}{2}}_{s=2}\\
|[(nn)^{I=1,\bar{3}_{c}}_{s=1}(ns)^{I=\frac{1}{2},\bar{3}_{c}}_{s=0}]^{I=\frac{3}{2},3_{c}}_{s=1}(cc)^{I=0,\bar{3}_{c}}_{s=1}\rangle^{I=\frac{3}{2}}_{s=2}
\end{pmatrix}$}
&
\multirow{4}*{$\begin{pmatrix}
5719\\5718\\5621\\5586
\end{pmatrix}$}
&1374.5&-1605.1&6.0&1.27&1.25&1.36&1.33&1.42&1.26&4.7&11.8&38.9&&&&55.4
\\
&&&1375.3&-1605.9&4.4&1.27&1.26&1.36&1.33&1.42&1.26&35.2&0.2&18.8&&&&54.2
\\
&&&1367.2&-1691.1&3.9&1.46&1.46&1.32&1.27&1.11&1.46&14.4&7.7&10.4&&&&32.5
\\
&&&1374.9&-1592.5&10.4&1.30&1.26&1.36&1.37&1.44&1.28&&&&&22.1&&22.1
\\
\toprule[0.05pt]
\multirow{6}*{$\frac{3}{2}(\frac{1}{2})[1^{+}]$}&
\multirow{6}*{$\begin{pmatrix}
|[(nn)^{I=1,\bar{3}_{c}}_{s=1}(ns)^{I=\frac{1}{2},\bar{3}_{c}}_{s=1}]^{I=\frac{3}{2},3_{c}}_{s=2}(cc)^{I=0,\bar{3}_{c}}_{s=1}\rangle^{I=\frac{3}{2}}_{s=1}\\
|[(nn)^{I=1,\bar{3}_{c}}_{s=1}(ns)^{I=\frac{1}{2},\bar{3}_{c}}_{s=1}]^{I=\frac{3}{2},3_{c}}_{s=1}(cc)^{I=0,\bar{3}_{c}}_{s=1}\rangle^{I=\frac{3}{2}}_{s=1}\\
|[(nn)^{I=1,\bar{3}_{c}}_{s=1}(ns)^{I=\frac{1}{2},\bar{3}_{c}}_{s=1}]^{I=\frac{3}{2},3_{c}}_{s=0}(cc)^{I=0,\bar{3}_{c}}_{s=1}\rangle^{I=\frac{3}{2}}_{s=1}\\
|[(nn)^{I=1,\bar{3}_{c}}_{s=1}(ns)^{I=\frac{1}{2},\bar{3}_{c}}_{s=1}]^{I=\frac{3}{2},\bar{6}_{c}}_{s=1}(cc)^{I=0,6_{c}}_{s=0}\rangle^{I=\frac{3}{2}}_{s=1}\\
|[(nn)^{I=1,\bar{3}_{c}}_{s=1}(ns)^{I=\frac{1}{2},\bar{3}_{c}}_{s=0}]^{I=\frac{3}{2},3_{c}}_{s=1}(cc)^{I=0,\bar{3}_{c}}_{s=1}\rangle^{I=\frac{3}{2}}_{s=1}\\
|[(nn)^{I=1,\bar{3}_{c}}_{s=1}(ns)^{I=\frac{1}{2},\bar{3}_{c}}_{s=0}]^{I=\frac{3}{2},\bar{6}_{c}}_{s=1}(cc)^{I=0,6_{c}}_{s=0}\rangle^{I=\frac{3}{2}}_{s=1}
\end{pmatrix}$}
&
\multirow{6}*{$\begin{pmatrix}
5710\\5708\\5707\\5629\\5575\\5488
\end{pmatrix}$}
&1379.9&-1610.5&-3.9&1.27&1.25&1.36&1.32&1.41&1.26&6.9&0.9&15.6&32.1&&&55.5
\\
&&&1380.7&-1611.2&-5.4&1.26&1.25&1.36&1.32&1.42&1.26&2.7&4.7&59.7&0.8&&&67.9
\\
&&&1381.1&-1611.6&-6.2&1.26&1.25&1.36&1.32&1.42&1.26&23.6&0.7&38.1&0.9&&&63.3
\\
&&&1364.7&-1688.5&8.5&1.47&1.46&1.32&1.28&1.11&1.46&23.4&2.5&3.4&3.5&&&32.8
\\
&&&1380.5&-1568.1&0.0&1.30&1.26&1.36&1.36&1.44&1.28&&&&&26.0&41.6&67.6
\\
&&&1368.7&-1649.4&6.1&1.49&1.47&1.31&1.32&1.13&1.48&&&&&21.0&7.7&28.7
\\
\toprule[0.05pt]
\multirow{3}*{$\frac{3}{2}(\frac{1}{2})[0^{+}]$}&
\multirow{3}*{$\begin{pmatrix}
|[(nn)^{I=1,\bar{3}_{c}}_{s=1}(ns)^{I=\frac{1}{2},\bar{3}_{c}}_{s=1}]^{I=\frac{3}{2},3_{c}}_{s=1}(cc)^{I=0,\bar{3}_{c}}_{s=1}\rangle^{I=\frac{3}{2}}_{s=0}\\
|[(nn)^{I=1,\bar{3}_{c}}_{s=1}(ns)^{I=\frac{1}{2},\bar{3}_{c}}_{s=1}]^{I=\frac{3}{2},\bar{6}_{c}}_{s=0}(cc)^{I=0,6_{c}}_{s=0}\rangle^{I=\frac{3}{2}}_{s=0}\\
|[(nn)^{I=1,\bar{3}_{c}}_{s=1}(ns)^{I=\frac{1}{2},\bar{3}_{c}}_{s=0}]^{I=\frac{3}{2},3_{c}}_{s=1}(cc)^{I=0,\bar{3}_{c}}_{s=1}\rangle^{I=\frac{3}{2}}_{s=0}
\end{pmatrix}$}
&
\multirow{3}*{$\begin{pmatrix}
5703\\5631\\5570
\end{pmatrix}$}
&1383.4&-1613.9&-10.3&1.26&1.25&1.35&1.32&1.41&1.26&28.2&&&20.0&&&48.2
\\
&&&1363.4&-1687.2&10.7&1.47&1.46&1.32&1.28&1.11&1.46&27.8&&&5.2&&&33.0
\\
&&&1383.4&-1570.9&-5.3&1.30&1.26&1.36&1.36&1.43&1.28&&&&&&97.5&97.5
\\
\bottomrule[1.00pt]
\multirow{2}*{$I[J^{P}]$}&\multirow{2}*{Configuration}&\multirow{2}*{Mass}&\multirow{2}*{$\langle T \rangle$}&\multirow{2}*{$\langle V^{\rm Con} \rangle$}&\multirow{2}*{$\langle V^{\rm SS} \rangle$}&\multirow{2}*{$R_{12}$}&\multirow{2}*{$R_{34}$}&$R_{13}$&$R_{23}$
&\multirow{2}*{$R_{12-34}$}
&$R_{13-24}$
&\multicolumn{2}{c}{\multirow{2}*{$\Lambda_{c}\Xi^{*}_{c}$}}
&\multicolumn{2}{c}{\multirow{2}*{$\Lambda_{c}\Xi'_{c}$}}
&\multicolumn{2}{c|}{\multirow{2}*{$\Lambda_{c}\Xi_{c}$}}
&\multicolumn{1}{c}{\multirow{2}*{$\Gamma_{sum}$}}\\
&&&&&&&&$R_{14}$&$R_{24}$&&$R_{14-23}$&&&&&&\\
\bottomrule[1.00pt]
\multirow{1}*{$\frac{1}{2}(2^{+})$}&$|[(nn)^{I=0,\bar{3}_{c}}_{s=0}(ns)^{I=\frac{1}{2},\bar{3}_{c}}_{s=1}]^{I=\frac{1}{2},3_{c}}_{s=1}(cc)^{I=0,\bar{3}_{c}}_{s=1}\rangle^{I=\frac{1}{2}}_{s=2}$
&5496&1380.9&-1506.8&9.7&1.35&1.26&1.45&1.33&1.46&1.30&\multicolumn{2}{c}{53.9}&\multicolumn{2}{c}{}&\multicolumn{2}{c|}{}&53.9
\\
\toprule[0.05pt]
\multirow{3}*{$\frac{1}{2}(1^{+})$}&
\multirow{3}*{$\begin{pmatrix}
|[(nn)^{I=0,\bar{3}_{c}}_{s=0}(ns)^{I=\frac{1}{2},\bar{3}_{c}}_{s=1}]^{I=\frac{1}{2},3_{c}}_{s=1}(cc)^{I=0,\bar{3}_{c}}_{s=1}\rangle^{I=\frac{1}{2}}_{s=1}\\
|[(nn)^{I=0,\bar{3}_{c}}_{s=0}(ns)^{I=\frac{1}{2},\bar{3}_{c}}_{s=1}]^{I=\frac{1}{2},\bar{6}_{c}}_{s=1}(cc)^{I=0,6_{c}}_{s=0}\rangle^{I=\frac{1}{2}}_{s=1}\\
|[(nn)^{I=0,\bar{3}_{c}}_{s=0}(ns)^{I=\frac{1}{2},\bar{3}_{c}}_{s=0}]^{I=\frac{1}{2},3_{c}}_{s=0}(cc)^{I=0,\bar{3}_{c}}_{s=1}\rangle^{I=\frac{1}{2}}_{s=1}
\end{pmatrix}$}
&
\multirow{3}*{$\begin{pmatrix}
5487\\5401\\5351
\end{pmatrix}$}
&1385.7&-1511.6&0.6&1.35&1.26&1.45&1.33&1.46&1.30&\multicolumn{2}{c}{10.3}&\multicolumn{2}{c}{6.1}&\multicolumn{2}{c|}{}&16.4
\\
&&&1376.5&-1593.9&6.1&1.54&1.48&1.43&1.26&1.14&1.50&\multicolumn{2}{c}{19.0}&\multicolumn{2}{c}{3.5}&\multicolumn{2}{c|}{}&22.5
\\
&&&1386.9&-1472.2&5.2&1.38&1.27&1.45&1.37&1.48&1.32&\multicolumn{2}{c}{}&\multicolumn{2}{c}{}&\multicolumn{2}{c|}{26.0}&26.0
\\
\toprule[0.05pt]
\multirow{2}*{$\frac{1}{2}(0^{+})$}&
\multirow{2}*{$\begin{pmatrix}
|[(nn)^{I=0,\bar{3}_{c}}_{s=0}(ns)^{I=\frac{1}{2},\bar{3}_{c}}_{s=1}]^{I=\frac{1}{2},3_{c}}_{s=1}(cc)^{I=0,\bar{3}_{c}}_{s=1}\rangle^{I=\frac{1}{2}}_{s=0}\\
|[(nn)^{I=0,\bar{3}_{c}}_{s=0}(ns)^{I=\frac{1}{2},\bar{3}_{c}}_{s=0}]^{I=\frac{3}{2},\bar{6}_{c}}_{s=0}(cc)^{I=0,6_{c}}_{s=0}\rangle^{I=\frac{1}{2}}_{s=0}
\end{pmatrix}$}
&
\multirow{2}*{$\begin{pmatrix}
5482\\5261
\end{pmatrix}$}
&1388.1&-1514.0&-4.0&1.35&1.26&1.45&1.33&1.46&1.30&\multicolumn{2}{c}{}&\multicolumn{2}{c}{4.7}&\multicolumn{2}{c|}{}&4.7
\\
&&&1378.7&-1555.0&6.0&1.56&1.49&1.42&1.30&1.16&1.53&\multicolumn{2}{c}{}&\multicolumn{2}{c}{}&\multicolumn{2}{c|}{18.5}&18.5
\\
\bottomrule[1.0pt]
\multicolumn{3}{l|}{$cc[ss][sn]$}&
\multicolumn{3}{c|}{Internal contribution}&
\multicolumn{6}{c|}{RMS Radius}
&\multicolumn{7}{c}{Fall-apart decay properties}\\
\Xcline{4-19}{0.3pt}
\multirow{2}*{$I(J^{P})$}&\multirow{2}*{Configuration}&\multirow{2}*{Mass}&\multirow{2}*{$\langle T \rangle$}&\multirow{2}*{$\langle V^{\rm Con} \rangle$}&\multirow{2}*{$\langle V^{\rm SS} \rangle$}&\multirow{2}*{$R_{12}$}&\multirow{2}*{$R_{34}$}&$R_{13}$&$R_{23}$
&\multirow{2}*{$R_{12-34}$}
&$R_{13-24}$
&\multirow{2}*{$\Omega^{*}_{c}\Xi^{*}_{c}$}&\multirow{2}*{$\Omega_{c}\Xi^{*}_{c}$}&\multirow{2}*{$\Omega^{*}_{c}\Xi'_{c}$}&\multirow{2}*{$\Omega_{c}\Xi_{c}$}&\multirow{2}*{$\Omega^{*}_{c}\Xi_{c}$}&\multirow{2}*{$\Omega_{c}\Xi_{c}$}&\multicolumn{1}{c}{\multirow{2}*{$\Gamma_{sum}$}}\\
 &&&&&&&&$R_{14}$&$R_{24}$&&$R_{14-23}$&&&&&&\\
\bottomrule[1.00pt]
\multirow{1}*{$\frac{1}{2}(3^{+})$}&$|[(ss)^{I=0,\bar{3}_{c}}_{s=1}(sn)^{I=\frac{1}{2},\bar{3}_{c}}_{s=1}]^{I=\frac{3}{2},3_{c}}_{s=2}(cc)^{I=0,\bar{3}_{c}}_{s=1}\rangle^{I=\frac{1}{2}}_{s=3}$&
5951&1364.5&-1659.1&18.3&1.22&1.25&1.31&1.33&1.39&1.24&51.1&&&&&&51.1
\\
\toprule[0.10pt]
\multirow{4}*{$\frac{1}{2}(2^{+})$}&
\multirow{4}*{$\begin{pmatrix}
|[(ss)^{I=0,\bar{3}_{c}}_{s=1}(sn)^{I=\frac{1}{2},\bar{3}_{c}}_{s=1}]^{I=\frac{3}{2},3_{c}}_{s=2}(cc)^{I=0,\bar{3}_{c}}_{s=1}\rangle^{I=\frac{1}{2}}_{s=2}\\
|[(ss)^{I=0,\bar{3}_{c}}_{s=1}(sn)^{I=\frac{1}{2},\bar{3}_{c}}_{s=1}]^{I=\frac{3}{2},3_{c}}_{s=1}(cc)^{I=0,\bar{3}_{c}}_{s=1}\rangle^{I=\frac{1}{2}}_{s=2}\\
|[(ss)^{I=0,\bar{3}_{c}}_{s=1}(sn)^{I=\frac{1}{2},\bar{3}_{c}}_{s=1}]^{I=\frac{3}{2},\bar{6}_{c}}_{s=2}(cc)^{I=0,6_{c}}_{s=0}\rangle^{I=\frac{1}{2}}_{s=2}\\
|[(ss)^{I=0,\bar{3}_{c}}_{s=1}(sn)^{I=\frac{1}{2},\bar{3}_{c}}_{s=0}]^{I=\frac{3}{2},3_{c}}_{s=1}(cc)^{I=0,\bar{3}_{c}}_{s=1}\rangle^{I=\frac{1}{2}}_{s=2}
\end{pmatrix}$}
&
\multirow{4}*{$\begin{pmatrix}
5938\\5937\\5844\\5805
\end{pmatrix}$}
&1371.6&-1666.2&5.3&1.22&1.25&1.30&1.33&1.39&1.23&3.8&43.6&15.2&&&&62.6
\\
&&&1371.7&-1666.3&5.1&1.22&1.25&1.30&1.33&1.39&1.23&39.8&20.7&0.8&&&&61.3
\\
&&&1363.3&-1750.7&4.5&1.42&1.44&1.25&1.29&1.09&1.43&15.8&13.0&9.7&&&&38.5
\\
&&&1371.7&-1621.6&9.4&1.25&1.25&1.31&1.37&1.41&1.25&&&&&27.7&&27.7
\\
\toprule[0.10pt]
\multirow{6}*{$\frac{1}{2}(1^{+})$}&
\multirow{6}*{$\begin{pmatrix}
|[(ss)^{I=0,\bar{3}_{c}}_{s=1}(sn)^{I=\frac{1}{2},\bar{3}_{c}}_{s=1}]^{I=\frac{3}{2},3_{c}}_{s=2}(cc)^{I=0,\bar{3}_{c}}_{s=1}\rangle^{I=\frac{1}{2}}_{s=1}\\
|[(ss)^{I=0,\bar{3}_{c}}_{s=1}(sn)^{I=\frac{1}{2},\bar{3}_{c}}_{s=1}]^{I=\frac{3}{2},3_{c}}_{s=1}(cc)^{I=0,\bar{3}_{c}}_{s=1}\rangle^{I=\frac{1}{2}}_{s=1}\\
|[(ss)^{I=0,\bar{3}_{c}}_{s=1}(sn)^{I=\frac{1}{2},\bar{3}_{c}}_{s=1}]^{I=\frac{3}{2},3_{c}}_{s=0}(cc)^{I=0,\bar{3}_{c}}_{s=1}\rangle^{I=\frac{1}{2}}_{s=1}\\
|[(ss)^{I=0,\bar{3}_{c}}_{s=1}(sn)^{I=\frac{1}{2},\bar{3}_{c}}_{s=1}]^{I=\frac{3}{2},\bar{6}_{c}}_{s=1}(cc)^{I=0,6_{c}}_{s=0}\rangle^{I=\frac{1}{2}}_{s=1}\\
|[(ss)^{I=0,\bar{3}_{c}}_{s=1}(sn)^{I=\frac{1}{2},\bar{3}_{c}}_{s=0}]^{I=\frac{3}{2},3_{c}}_{s=1}(cc)^{I=0,\bar{3}_{c}}_{s=1}\rangle^{I=\frac{1}{2}}_{s=1}\\
|[(ss)^{I=0,\bar{3}_{c}}_{s=1}(sn)^{I=\frac{1}{2},\bar{3}_{c}}_{s=0}]^{I=\frac{3}{2},\bar{6}_{c}}_{s=1}(cc)^{I=0,6_{c}}_{s=0}\rangle^{I=\frac{1}{2}}_{s=1}
\end{pmatrix}$}
&
\multirow{6}*{$\begin{pmatrix}
5930\\5929\\5928\\5848\\5797\\5709
\end{pmatrix}$}
&1376.4&-1671.0&-3.4&1.22&1.25&1.30&1.32&1.38&1.23&6.6&15.4&1.3&39.6&&&62.9
\\
&&&1376.5&-1671.0&-3.6&1.22&1.25&1.30&1.32&1.39&1.23&2.2&62.7&7.5&0.7&&&73.1
\\
&&&1376.5&-1671.0&-3.7&1.21&1.25&1.30&1.32&1.39&1.23&25.2&40.4&2.1&1.4&&&69.1
\\
&&&1361.3&-1748.7&8.1&1.42&1.44&1.25&1.29&1.09&1.43&25.9&4.3&3.2&5.3&&&38.7
\\
&&&1376.2&-1626.0&1.1&1.25&1.25&1.30&1.36&1.41&1.25&&&&&26.1&45.5&71.6
\\
&&&1365.3&-1708.2&6.2&1.45&1.45&1.25&1.33&1.10&1.45&&&&&23.6&9.3&32.9
\\
\toprule[0.10pt]
\multirow{3}*{$\frac{1}{2}(0^{+})$}&
\multirow{3}*{$\begin{pmatrix}
|[(ss)^{I=0,\bar{3}_{c}}_{s=1}(sn)^{I=\frac{1}{2},\bar{3}_{c}}_{s=1}]^{I=\frac{3}{2},3_{c}}_{s=1}(cc)^{I=0,\bar{3}_{c}}_{s=1}\rangle^{I=\frac{1}{2}}_{s=0}\\
|[(ss)^{I=0,\bar{3}_{c}}_{s=1}(sn)^{I=\frac{1}{2},\bar{3}_{c}}_{s=1}]^{I=\frac{3}{2},\bar{6}_{c}}_{s=0}(cc)^{I=0,6_{c}}_{s=0}\rangle^{I=\frac{1}{2}}_{s=0}\\
|[(ss)^{I=0,\bar{3}_{c}}_{s=1}(sn)^{I=\frac{1}{2},\bar{3}_{c}}_{s=0}]^{I=\frac{3}{2},3_{c}}_{s=1}(cc)^{I=0,\bar{3}_{c}}_{s=1}\rangle^{I=\frac{1}{2}}_{s=0}
\end{pmatrix}$}
&
\multirow{3}*{$\begin{pmatrix}
5925\\5850\\5793
\end{pmatrix}$}
&1378.8&-1673.4&-8.0&1.21&1.25&1.30&1.32&1.39&1.23&27.7&&&32.7&&&60.4
\\
&&&1360.4&-1747.7&9.8&1.42&1.44&1.26&1.29&1.09&1.43&30.8&&&7.8&&&38.6
\\
&&&1378.4&-1628.3&-3.1&1.25&1.25&1.30&1.36&1.41&1.25&&&&&&99.7&99.7
\\
\bottomrule[0.50pt]
\bottomrule[1.50pt]
\end{tabular}
\end{lrbox}\scalebox{0.79}{\usebox{\tablebox}}
\end{table*}

\begin{figure*}[htbp]
\begin{tabular}{c}
\includegraphics[width=\textwidth]{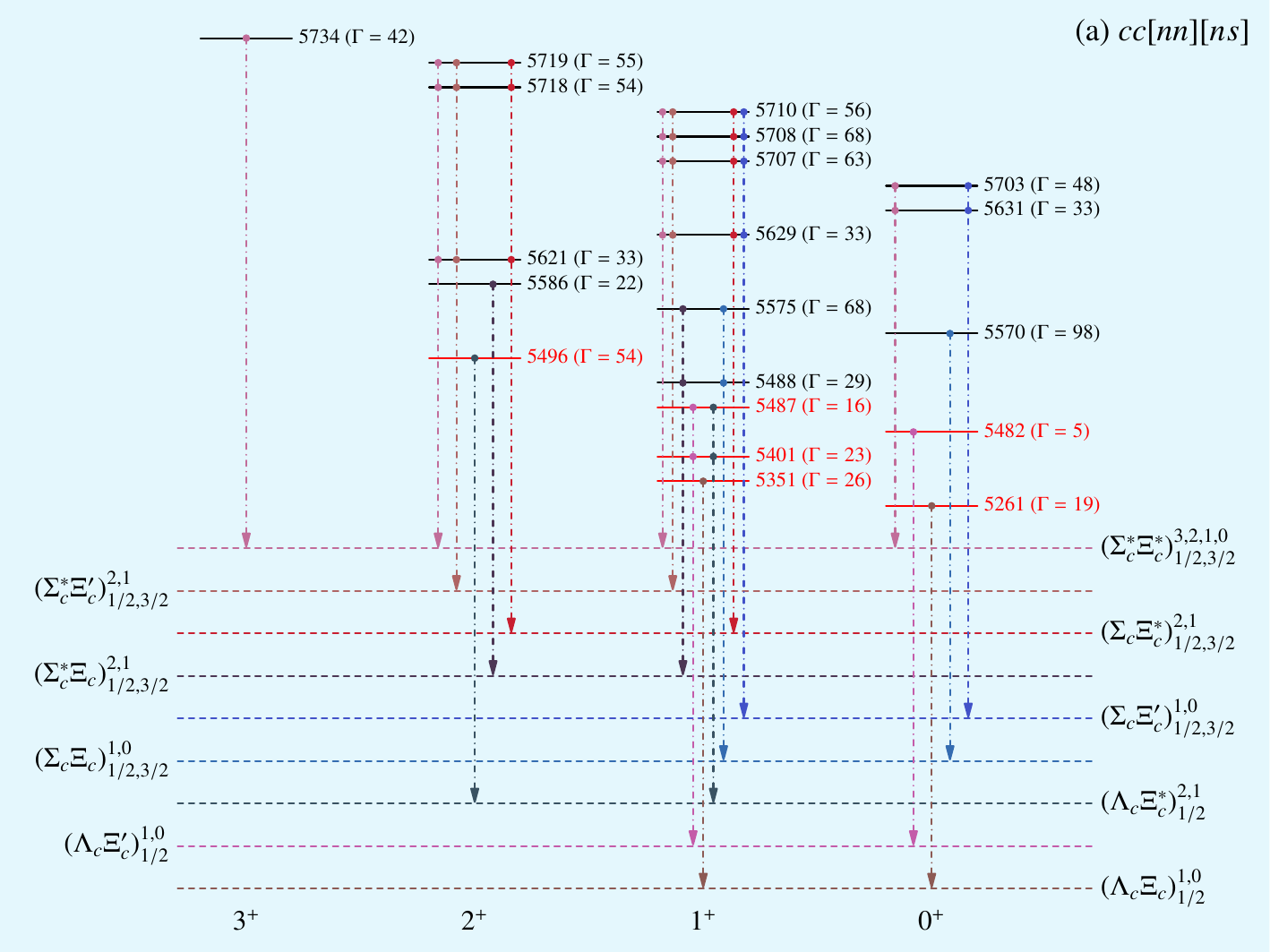}\\
\includegraphics[width=\textwidth]{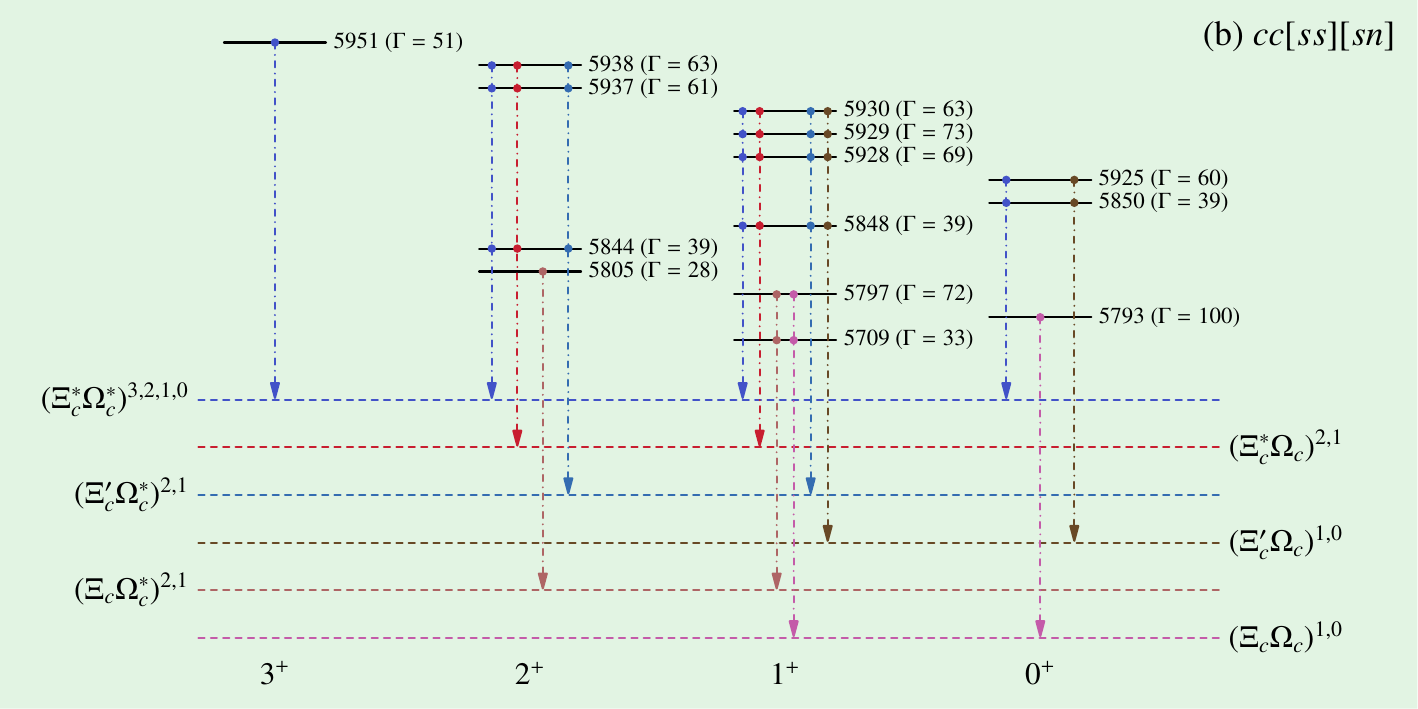}\\
\end{tabular}
\caption{
Relative positions for the $cc[nn][ns]$ (a) and $cc[ss][sn]$ (b) hexaquark states labeled with horizontal solid lines, the labels, e.g. $5703~(\Gamma=48)$ represents the mass and total decay width of the corresponding state (units: MeV).
In the $cc[nn][ns]$ subsystem, the black and red horizontal lines represent the hexaquark states with $I=3/2$ and $I=1/2$, respectively.
The notations are same as those of Fig. \ref{fig-ccnnnn}.
}\label{fig-ccnnns}
\end{figure*}

Next, we discuss the $cc[nn][ns]$ and $cc[ss][sn]$ subsystems.
For the  $cc[nn][ns]$ subsystem, $|[(nn)^{I=0}(ns)^{I=\frac{1}{2}}]\\(cc)\rangle$ configuration only couples to $I=1/2$ states.
On the contrary, $|[(nn)^{I=1}(ns)^{I=\frac{1}{2}}](cc)\rangle$ configuration can couple to $I=3/2$ and $I=1/2$ states.
Although color-spin mixing occurs in two above configuration for the $I=1/2$ states, as can be seen from Table \ref{ccnnns}, the $\langle V^{\rm SS}\rangle$ is significantly suppressed, resulting in the mixing negligible.
Thus, the $I=3/2$ and $I=1/2$ states from the $|[(nn)^{I=1}(ns)^{I=\frac{1}{2}}](cc)\rangle$ configuration would have the same mass spectra, RMS radii, and decay behaviors.
For these states, they are relatively broad states, and have many different decay channels.
Therefore, there are obstacles to experimentally discovering these states. Their resonance peaks are prone to be misjudged as continuous background fluctuations. Thus, we do not recommend that relevant experiments attempt to reconstruct resonance peaks from these decay final states.

Here, we take the 
${\rm H}_{c^{2}[nn][ns]}(5710, 3/2(1/2), 2^{+})$ as an example for discussion. 
Similar situations also apply to other states.
Its total decay width reaches 56 MeV and relative partial decay width ratio is:
\begin{eqnarray}
\Gamma_{\Sigma^{*}_{c}\Xi^{*}_{c}}:\Gamma_{\Sigma^{*}_{c}\Xi'_{c}}:\Gamma_{\Sigma_{c}\Xi^{*}_{c}}:\Gamma_{\Sigma_{c}\Xi'_{c}}=7.7:1:17:36,
\end{eqnarray}
i.e. $\Sigma_{c}\Xi^{*}_{c}$ and $\Sigma_{c}\Xi'_{c}$ are its dominant decay channels.

According to Fig.~\ref{fig-ccnnns}, for the $cc[nn][ns]$ states with the configuration 
$|[(nn)^{I = 0}(ns)^{I=\frac{1}{2}}](cc)\rangle$, 
their masses are generally lower than those of the states with the configuration 
$|[(nn)^{I = 1}(ns)^{I=\frac{1}{2}}](cc)\rangle$.
This is because the mass of the scalar diquark $[nn]$ is lower than that of the vector diquark $[nn]$ with $I = 1$ and $S = 1$. Meanwhile, the total decay widths of $|[(nn)^{I = 0}(ns)^{I=\frac{1}{2}}](cc)\rangle$  are generally smaller than those of $|[(nn)^{I = 1}(ns)^{I=\frac{1}{2}}](cc)\rangle$, and the same situation exists in the $cc[nn][nn]$ and $cc[nn][ss]$ subsystems. This indicates that compared with other states, those containing the scalar diquark configuration have lower masses, stronger internal interactions, narrower total widths, relatively longer lifetimes, and are more likely to be detected experimentally in the doubly charmed hexaquark system.

Moreover,
there exists a relatively narrow state, ${\rm H}_{c^{2}[nn][ns]}(5482, 1/2, 0^{+})$, whose total width is less than 5 MeV and 
it only decays to $\Lambda_{c}\Xi'_{c}$ final states.
Its relatively narrow width implies that its resonance peak is quite distinct, and the specific decay mode provides a clear signature for experimental searches. 
Therefore, the ${\rm H}_{c^{2}[nn][ns]}(5482, 1/2, 0^{+})$ has the potential to be discovered in experiments. 
Further, we suggest the relevant experiment check for the signal of ${\rm H}_{c^{2}[nn][ns]}(5482, 1/2, 0^{+})$ in the $5400$-$5500$ MeV energy window, 
and its lineshape should be relatively prominent in $\Lambda_{c}\Xi'_{c}$ mass spectrum.

In addition, the ${\rm H}_{c^{2}[nn][ns]}(5487, 1/2, 1^{+})$ and ${\rm H}_{c^{2}[nn][ns]}(5261, 1/2, 0^{+})$ are also relatively narrow states, with total widths of 16 and 19 MeV, respectively.
The ${\rm H}_{c^{2}[nn][ns]}(5261, 1/2, 0^{+})$ only decays to $\Xi_{c}\Lambda_{c}$ final states.
For the ${\rm H}_{c^{2}[nn][ns]}(5487, 1/2, 1^{+})$, we obtain its relative partial decay width ratios as:
\begin{eqnarray}
\frac{\Gamma[{\rm H}_{c^{2}[nn][ns]}(5487, 1/2, 1^{+})\to \Xi^{*}_{c}\Lambda_{c}]}{\Gamma[{\rm H}_{c^{2}[nn][ns]}(5487, 1/2, 1^{+})\to\Xi'_{c}\Lambda_{c}]}=\frac{5}{3}.
\end{eqnarray}

For the $cc[ss][sn]$ subsystem, since it has exactly the same symmetry constraints as $cc[nn][ns]$  subsystem with $I=3/2$, the number of allowed states is also exactly identical. 
Due to the larger decay phase space and multiple different decay channels, 
most of the states belong to relatively broad states, with their decay widths ranging from 35 to 100 MeV.
Among them,  there is a relatively special state: ${\rm H}_{c^{2}[ss][sn]}(5806, 1/2, 2^{+})$, whose total width is 27 MeV.
Therefore, it is the narrowest state in the $cc[ss][sn]$ subsystem. 
Moreover, it only decays into the $\Omega^{*}_{c}\Xi_{c}$ final states, which means there is a high possibility of observing it in the $\Omega^{*}_{c}\Xi_{c}$ decay channel.
Based on the above research on the typical states, and with reference to Table \ref{ccnnns} and Fig. \ref{fig-ccnnns}, one can perform similar discussions on the decay behaviors of other $cc[ss][sn]$ states, and further explore their characteristics in-depth.
\\

\subsection{The $cc[nn][ss]$ and $cc[ns][ns]$ subsystems}

\begin{figure*}[htbp]
\begin{tabular}{c}
\includegraphics[width=\textwidth]{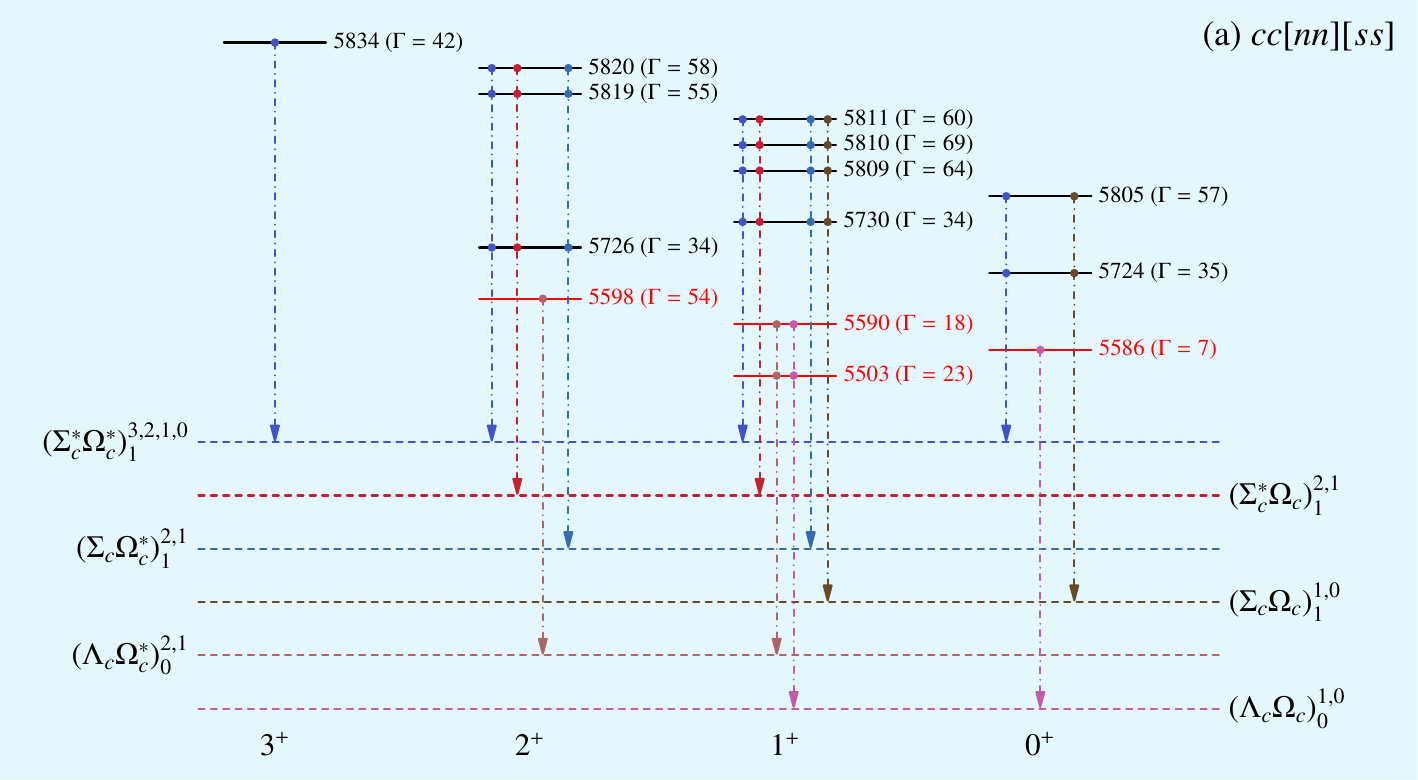}\\
\includegraphics[width=\textwidth]{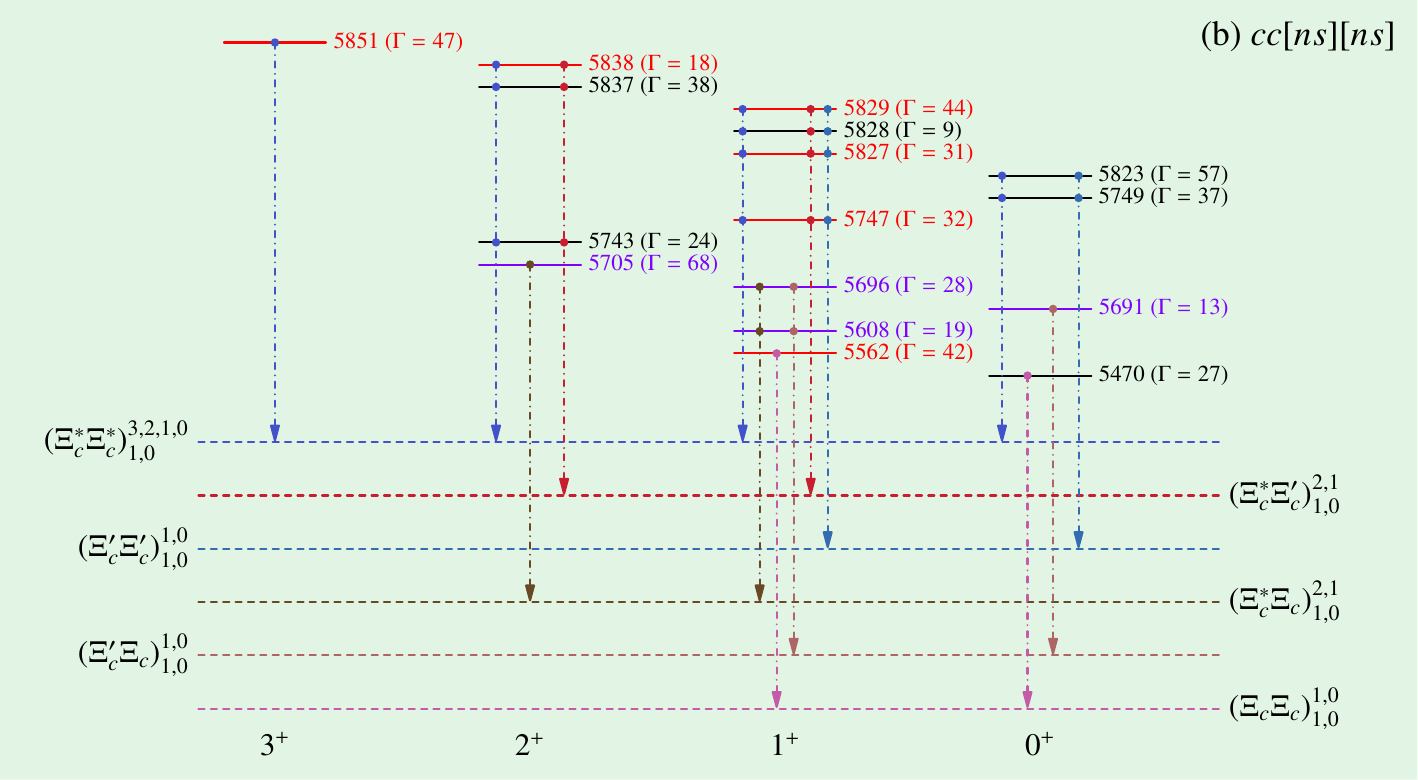}\\
\end{tabular}
\caption{
Relative positions for the 
$cc[nn][ss]$ (a) and $cc[ns][ns]$ (b) hexaquark states labeled with horizontal solid lines, the labels, e.g. $5703~(\Gamma=48)$ represents the mass and total decay width of the corresponding state (units: MeV).
In the $cc[nn][ss]$ and $cc[ns][ns]$ subsystems, the black, red, and purple horizontal lines represent the hexaquark states with $I=1$, $I=0$, and $I=1,0$, respectively.
The notations are same as those of Fig. \ref{fig-ccnnnn}.
}\label{fig-ccnnss}
\end{figure*}

\begin{table*}[t]
\centering \caption{
The numerical results of the mass spectrum, the mass contributions of each Hamiltonian part (in MeV), the root-mean-square radii (in fm), and the partial decay widths and total decay widths of the fall-apart decay processes (in MeV) for the $cc[nn][ss]$ and $cc[ns][ns]$ hexaquark states.
}\label{ccnnss}
\begin{lrbox}{\tablebox}
\renewcommand\arraystretch{1.8}
\renewcommand\tabcolsep{1.2 pt}
\begin{tabular}{ccc|ccc|cccccc|rrrr|r}
\toprule[1.50pt]
\toprule[0.50pt]
\multicolumn{3}{l|}{$cc[nn][ss]$}&
\multicolumn{3}{c|}{Each part contribution}& 
\multicolumn{6}{c|}{RMS Radius}&\multicolumn{5}{c}{Fall-apart decay properties}\\
\Xcline{4-17}{0.3pt}
\multirow{2}*{$I[J^{P}]$}&\multirow{2}*{Configuration}&\multirow{2}*{Mass}&\multirow{2}*{$\langle T \rangle$}
&\multirow{2}*{$\langle V^{\rm Con} \rangle$}
&\multirow{2}*{$\langle V^{\rm SS} \rangle$}
&\multirow{2}*{$R_{12}$}&\multirow{2}*{$R_{34}$}
&$R_{13}$&$R_{23}$
&\multirow{2}*{$R_{12-34}$}
&$R_{13-24}$
&\multirow{2}*{$\Sigma^{*}_{c}\Omega^{*}_{c}$}
&\multirow{2}*{$\Sigma^{*}_{c}\Omega_{c}$}
&\multirow{2}*{$\Sigma_{c}\Omega^{*}_{c}$}
&\multirow{2}*{$\Sigma_{c}\Omega_{c}$}
&\multicolumn{1}{c}{\multirow{2}*{$\Gamma_{sum}$}}
\\
&&&&&&&&$R_{14}$&$R_{24}$&&$R_{14-23}$&&&&&\\
\bottomrule[1.00pt]
\multirow{1}*{$1[3^{+}]$}&$|[(nn)^{I=1,\bar{3}_{c}}_{s=1}(ss)^{I=0,\bar{3}_{c}}_{s=1}]^{I=1,3_{c}}_{s=2}(cc)^{I=0,\bar{3}_{c}}_{s=1}\rangle^{I=1}_{s=3}$&
5834&1365.7&-1622.2&19.7&1.25&1.25&1.36&1.31&1.25&1.41&42.2&&&&42.2
\\
\toprule[0.10pt]
\multirow{3}*{$1[2^{+}]$}&
\multirow{3}*{$\begin{pmatrix}
|[(nn)^{I=1,\bar{3}_{c}}_{s=1}(ss)^{I=0,\bar{3}_{c}}_{s=1}]^{I=1,3_{c}}_{s=2}(cc)^{I=0,\bar{3}_{c}}_{s=1}\rangle^{I=1}_{s=2}\\
|[(nn)^{I=1,\bar{3}_{c}}_{s=1}(ss)^{I=0,\bar{3}_{c}}_{s=1}]^{I=1,3_{c}}_{s=1}(cc)^{I=0,\bar{3}_{c}}_{s=1}\rangle^{I=1}_{s=2}\\
|[(nn)^{I=1,\bar{3}_{c}}_{s=1}(ss)^{I=0,\bar{3}_{c}}_{s=1}]^{I=1,\bar{6}_{c}}_{s=2}(cc)^{I=0,6_{c}}_{s=0}\rangle^{I=1}_{s=2}
\end{pmatrix}$}
&
\multirow{3}*{$\begin{pmatrix}
5820\\5819\\5726
\end{pmatrix}$}
&1373.4&-1629.9&5.7&1.25&1.25&1.36&1.30&1.40&1.25&4.5&13.4&39.7&&57.6
\\
&&&1373.8&-1630.3&4.8&1.25&1.25&1.36&1.30&1.41&1.25&35.3&0.4&18.8&&54.5
\\
&&&1366.0&-1715.5&4.2&1.44&1.45&1.32&1.25&1.10&1.44&14.3&9.3&10.7&&34.3
\\
\toprule[0.10pt]
\multirow{4}*{$1[1^{+}]$}&
\multirow{4}*{$\begin{pmatrix}
|[(nn)^{I=1,\bar{3}_{c}}_{s=1}(ss)^{I=0,\bar{3}_{c}}_{s=1}]^{I=1,3_{c}}_{s=2}(cc)^{I=0,\bar{3}_{c}}_{s=1}\rangle^{I=1}_{s=1}\\
|[(nn)^{I=1,\bar{3}_{c}}_{s=1}(ss)^{I=0,\bar{3}_{c}}_{s=1}]^{I=1,3_{c}}_{s=1}(cc)^{I=0,\bar{3}_{c}}_{s=1}\rangle^{I=1}_{s=1}\\
|[(nn)^{I=1,\bar{3}_{c}}_{s=1}(ss)^{I=0,\bar{3}_{c}}_{s=1}]^{I=1,3_{c}}_{s=0}(cc)^{I=0,\bar{3}_{c}}_{s=1}\rangle^{I=1}_{s=1}\\
|[(nn)^{I=1,\bar{3}_{c}}_{s=1}(ss)^{I=0,\bar{3}_{c}}_{s=1}]^{I=1,\bar{6}_{c}}_{s=1}(cc)^{I=0,6_{c}}_{s=0}\rangle^{I=1}_{s=1}
\end{pmatrix}$}
&
\multirow{4}*{$\begin{pmatrix}
5811\\5810\\5809\\5730
\end{pmatrix}$}
&1378.5&-1635.0&-3.7&1.25&1.25&1.35&1.30&1.40&1.25&7.0&1.1&15.7&35.9&59.7
\\
&&&1378.9&-1635.4&-4.6&1.24&1.25&1.36&1.30&1.40&1.25&2.6&5.7&59.8&0.8&68.9
\\
&&&1379.2&-1635.6&-5.1&1.24&1.25&1.36&1.30&1.41&1.25&23.7&1.1&38.1&1.1&64.0
\\
&&&1363.7&-1713.2&8.3&1.45&1.45&1.32&1.25&1.10&1.45&23.4&2.7&3.5&4.0&33.6
\\
\toprule[0.10pt]
\multirow{2}*{$1[0^{+}]$}&
\multirow{2}*{$\begin{pmatrix}
|[(nn)^{I=1,\bar{3}_{c}}_{s=1}(ss)^{I=0,\bar{3}_{c}}_{s=1}]^{I=1,3_{c}}_{s=1}(cc)^{I=0,\bar{3}_{c}}_{s=1}\rangle^{I=1}_{s=0}\\
|[(nn)^{I=1,\bar{3}_{c}}_{s=1}(ss)^{I=0,\bar{3}_{c}}_{s=1}]^{I=1,\bar{6}_{c}}_{s=0}(cc)^{I=0,6_{c}}_{s=0}\rangle^{I=1}_{s=0}
\end{pmatrix}$}
&
\multirow{2}*{$\begin{pmatrix}
5805\\5724
\end{pmatrix}$}
&1381.5&-1638.0&-9.3&1.24&1.25&1.35&1.30&1.40&1.25&28.5&&&28.2&
56.7
\\
&&&1367.1&-1714.5&2.1&1.44&1.45&1.32&1.25&1.10&1.44&28.9&&&6.2&35.1
\\
\bottomrule[1.00pt]
\multirow{2}*{$I[J^{P}]$}&\multirow{2}*{Configuration}&\multirow{2}*{Mass}&\multirow{2}*{$\langle T \rangle$}
&\multirow{2}*{$\langle V^{\rm Con} \rangle$}
&\multirow{2}*{$\langle V^{\rm SS} \rangle$}
&\multirow{2}*{$R_{12}$}&\multirow{2}*{$R_{34}$}
&$R_{13}$&$R_{23}$
&\multirow{2}*{$R_{12-34}$}
&$R_{13-24}$
&\multicolumn{2}{c}{\multirow{2}*{$\Lambda_{c}\Omega^{*}_{c}$}}
&\multicolumn{2}{c|}{\multirow{2}*{$\Lambda_{c}\Omega_{c}$}}
&\multicolumn{1}{c}{\multirow{2}*{$\Gamma_{sum}$}}\\
&&&&&&&&$R_{14}$&$R_{24}$&&$R_{14-23}$&&&&&\\
\bottomrule[1.00pt]
\multirow{1}*{$0[2^{+}]$}&$|[(nn)^{I=0,\bar{3}_{c}}_{s=0}(ss)^{I=0,\bar{3}_{c}}_{s=1}]^{I=0,3_{c}}_{s=1}(cc)^{I=0,\bar{3}_{c}}_{s=1}\rangle^{I=0}_{s=2}$&
5598&1379.3&-1529.5&9.3&1.33&1.26&1.45&1.31&1.45&1.29&\multicolumn{2}{c}{53.6}&\multicolumn{2}{c|}{}&53.6
\\
\toprule[0.10pt]
\multirow{2}*{$0[1^{+}]$}&
\multirow{2}*{$\begin{pmatrix}
|[(nn)^{I=0,\bar{3}_{c}}_{s=0}(ss)^{I=0,\bar{3}_{c}}_{s=1}]^{I=0,3_{c}}_{s=1}(cc)^{I=0,\bar{3}_{c}}_{s=1}\rangle^{I=0}_{s=1}\\
|[(nn)^{I=0,\bar{3}_{c}}_{s=0}(ss)^{I=0,\bar{3}_{c}}_{s=1}]^{I=1,\bar{6}_{c}}_{s=1}(cc)^{I=0,6_{c}}_{s=0}\rangle^{I=0}_{s=1}
\end{pmatrix}$}
&
\multirow{2}*{$\begin{pmatrix}
5590\\5503
\end{pmatrix}$}
&1383.6&-1533.9&1.1&1.33&1.26&1.45&1.31&1.44&1.29&\multicolumn{2}{c}{10.6}&\multicolumn{2}{c|}{7.5}&18.1
\\
&&&1375.7&-1617.5&6.1&1.52&1.47&1.43&1.24&1.13&1.48&\multicolumn{2}{c}{18.8}&\multicolumn{2}{c|}{4.0}&22.8
\\
\toprule[0.10pt]
\multirow{1}*{$0[0^{+}]$}&$|[(nn)^{I=0,\bar{3}_{c}}_{s=0}(ss)^{I=0,\bar{3}_{c}}_{s=1}]^{I=0,3_{c}}_{s=1}(cc)^{I=0,\bar{3}_{c}}_{s=1}\rangle^{I=0}_{s=0}$&
5586&1385.8&-1536.1&-3.1&1.33&1.26&1.45&1.30&1.44&1.29&\multicolumn{2}{c}{}&\multicolumn{2}{c|}{6.5}&6.5
\\
\bottomrule[1.00pt]
\multicolumn{3}{l|}{$cc[ns][ns]$}&
\multicolumn{3}{c|}{Each part contribution}& \multicolumn{6}{c|}{RMS Radius}&\multicolumn{5}{c}{Fall-apart decay properties}\\
\Xcline{4-17}{0.3pt}
\multirow{2}*{$I[J^{P}]$}&\multirow{2}*{Configuration}&\multirow{2}*{Mass}&\multirow{2}*{$\langle T \rangle$}&\multirow{2}*{$\langle V^{\rm Con} \rangle$}&\multirow{2}*{$\langle V^{\rm SS} \rangle$}&\multirow{2}*{$R_{12}$}&\multirow{2}*{$R_{34}$}&$R_{13}$&$R_{23}$
&\multirow{2}*{$R_{12-34}$}
&$R_{13-24}$
&\multirow{2}*{$\Xi^{*}_{c}\Xi^{*}_{c}$}&\multirow{2}*{$\Xi^{*}_{c}\Xi'_{c}$}&\multirow{2}*{$\Xi'_{c}\Xi'_{c}$}&\multirow{2}*{$\Xi_{c}\Xi_{c}$}&\multicolumn{1}{c}{\multirow{2}*{$\Gamma_{sum}$}}\\
 &&&&&&&&$R_{14}$&$R_{24}$&&$R_{14-23}$&&&&&\\
\bottomrule[1.00pt]
\multirow{2}*{$1[2^{+}]$}&
\multirow{2}*{$\begin{pmatrix}
|[(ns)^{I=\frac{1}{2},\bar{3}_{c}}_{s=1}(ns)^{I=\frac{1}{2},\bar{3}_{c}}_{s=1}]^{I=1,3_{c}}_{s=1}(cc)^{I=0,\bar{3}_{c}}_{s=1}\rangle^{I=1}_{s=2}\\
|[(ns)^{I=\frac{1}{2},\bar{3}_{c}}_{s=1}(ns)^{I=\frac{1}{2},\bar{3}_{c}}_{s=1}]^{I=1,\bar{6}_{c}}_{s=2}(cc)^{I=0,6_{c}}_{s=0}\rangle^{I=1}_{s=2}
\end{pmatrix}$}
&
\multirow{2}*{$\begin{pmatrix}
5837\\5743
\end{pmatrix}$}
&1373.0&-1641.0&4.8&1.24&1.25&1.33&1.33&1.40&1.24&37.6&0.4&&&38.0
\\
&&&1364.7&-1725.8&4.3&1.44&1.45&1.28&1.28&1.10&1.44&15.3&8.9&&&24.2
\\
\toprule[0.10pt]
\multirow{1}*{$1(1^{+})$}&$|[(ns)^{I=\frac{1}{2},\bar{3}_{c}}_{s=1}(ns)^{I=\frac{1}{2},\bar{3}_{c}}_{s=1}]^{I=1,3_{c}}_{s=1}(cc)^{I=0,\bar{3}_{c}}_{s=1}\rangle^{I=1}_{s=1}$&
5828&1378.0&-1646.0&-4.3&1.24&1.25&1.32&1.32&1.40&1.24&2.4&6.1&0.7&&9.2
\\
\toprule[0.10pt]
\multirow{3}*{$1[0^{+}]$}&
\multirow{3}*{$\begin{pmatrix}
|[(ns)^{I=\frac{1}{2},\bar{3}_{c}}_{s=1}(ns)^{I=\frac{1}{2},\bar{3}_{c}}_{s=1}]^{I=1,3_{c}}_{s=1}(cc)^{I=0,\bar{3}_{c}}_{s=1}\rangle^{I=1}_{s=0}\\
|[(ns)^{I=\frac{1}{2},\bar{3}_{c}}_{s=1}(ns)^{I=\frac{1}{2},\bar{3}_{c}}_{s=1}]^{I=1,\bar{6}_{c}}_{s=0}(cc)^{I=0,6_{c}}_{s=0}\rangle^{I=1}_{s=0}\\
|[(ns)^{I=\frac{1}{2},\bar{3}_{c}}_{s=0}(ns)^{I=\frac{1}{2},\bar{3}_{c}}_{s=0}]^{I=1,\bar{6}_{c}}_{s=0}(cc)^{I=0,6_{c}}_{s=0}\rangle^{I=1}_{s=0}
\end{pmatrix}$}
&
\multirow{3}*{$\begin{pmatrix}
5823\\5749\\5470
\end{pmatrix}$}
&1380.5&-1648.5&-8.9&1.23&1.25&1.32&1.32&1.40&1.24&27.6&&29.5&&57.1
\\
&&&1361.4&-1722.5&10.2&1.44&1.45&1.28&1.28&1.10&1.44&30.0&&6.7&&36.7
\\
&&&1369.1&-1643.4&6.1&1.50&1.47&1.32&1.32&1.13&1.48&&&&26.7&26.7
\\
\bottomrule[1.00pt]
\multirow{1}*{$0[3^{+}]$}&$|[(ns)^{I=\frac{1}{2},\bar{3}_{c}}_{s=1}(ns)^{I=\frac{1}{2},\bar{3}_{c}}_{s=1}]^{I=0,3_{c}}_{s=2}(cc)^{I=0,\bar{3}_{c}}_{s=1}\rangle^{I=0}_{s=3}$&
5851&1365.2&-1633.2&19.3&1.24&1.25&1.33&1.33&1.41&1.25&47.1&&&&47.1
\\
\toprule[0.10pt]
\multirow{1}*{$0[2^{+}]$}&
$|[(ns)^{I=\frac{1}{2},\bar{3}_{c}}_{s=1}(ns)^{I=\frac{1}{2},\bar{3}_{c}}_{s=1}]^{I=0,3_{c}}_{s=2}(cc)^{I=0,\bar{3}_{c}}_{s=1}\rangle^{I=0}_{s=2}$
&
5838&1372.7&-1640.7&5.6&1.24&1.25&1.33&1.33&1.40&1.25&4.2&13.4&&&17.6
\\
\toprule[0.10pt]
\multirow{4}*{$0[1^{+}]$}&
\multirow{4}*{$\begin{pmatrix}
|[(ns)^{I=\frac{1}{2},\bar{3}_{c}}_{s=1}(ns)^{I=\frac{1}{2},\bar{3}_{c}}_{s=1}]^{I=0,3_{c}}_{s=2}(cc)^{I=0,\bar{3}_{c}}_{s=1}\rangle^{I=0}_{s=1}\\
|[(ns)^{I=\frac{1}{2},\bar{3}_{c}}_{s=1}(ns)^{I=\frac{1}{2},\bar{3}_{c}}_{s=1}]^{I=0,3_{c}}_{s=0}(cc)^{I=0,\bar{3}_{c}}_{s=1}\rangle^{I=0}_{s=1}\\
|[(ns)^{I=\frac{1}{2},\bar{3}_{c}}_{s=1}(ns)^{I=\frac{1}{2},\bar{3}_{c}}_{s=1}]^{I=0,\bar{6}_{c}}_{s=1}(cc)^{I=0,6_{c}}_{s=0}\rangle^{I=0}_{s=1}\\
|[(ns)^{I=\frac{1}{2},\bar{3}_{c}}_{s=0}(ns)^{I=\frac{1}{2},\bar{3}_{c}}_{s=0}]^{I=0,3_{c}}_{s=0}(cc)^{I=0,\bar{3}_{c}}_{s=1}\rangle^{I=0}_{s=1}
\end{pmatrix}$}
&
\multirow{4}*{$\begin{pmatrix}
5829\\5827\\5747\\5562
\end{pmatrix}$}
&1377.7&-1645.7&-3.6&1.24&1.25&1.32&1.32&1.40&1.24&6.6&1.1&36.4&&44.1
\\
&&&1378.2&-1646.2&-4.7&1.23&1.25&1.32&1.32&1.40&1.24&28.6&1.4&1.3&&31.3
\\
&&&1362.5&-1723.6&8.2&1.44&1.45&1.28&1.28&1.10&1.44&25.0&2.9&4.5&&32.4
\\
&&&1378.2&-1559.3&5.2&1.30&1.26&1.36&1.36&1.44&1.28&&&&41.5&41.5
\\
\bottomrule[1.00pt]
\multirow{2}*{$I(J^{P})$}&\multirow{2}*{Configuration}&\multirow{2}*{Mass}&\multirow{2}*{$\langle T \rangle$}
&\multirow{2}*{$\langle V^{\rm Con} \rangle$}
&\multirow{2}*{$\langle V^{\rm SS} \rangle$}
&\multirow{2}*{$R_{12}$}&\multirow{2}*{$R_{34}$}
&$R_{13}$&$R_{23}$
&\multirow{2}*{$R_{12-34}$}
&$R_{13-24}$
&\multicolumn{2}{c}{\multirow{2}*{$\Xi_{c}\Xi^{*}_{c}$}}
&\multicolumn{2}{c|}{\multirow{2}*{$\Xi_{c}\Xi'_{c}$}}
&\multicolumn{1}{c}{\multirow{2}*{$\Gamma_{sum}$}}\\
&&&&&&&&$R_{14}$&$R_{24}$&&$R_{14-23}$&&&&&\\
\bottomrule[1.00pt]
\multirow{1}*{$1(0)[2^{+}]$}&$|[(ns)^{I=\frac{1}{2},\bar{3}_{c}}_{s=0}(ns)^{I=\frac{1}{2},\bar{3}_{c}}_{s=1}]^{I=1,3_{c}}_{s=1}(cc)^{I=0,\bar{3}_{c}}_{s=1}\rangle^{I=1}_{s=2}$&
5705&1372.9&-1596.9&9.8&1.27&1.26&1.37&1.33&1.42&1.26&\multicolumn{2}{c}{68.4}&\multicolumn{2}{c|}{}&68.4
\\
\toprule[0.10pt]
\multirow{2}*{$1(0)[1^{+}]$}&
\multirow{2}*{$\begin{pmatrix}
|[(ns)^{I=\frac{1}{2},\bar{3}_{c}}_{s=0}(ns)^{I=\frac{1}{2},\bar{3}_{c}}_{s=1}]^{I=1,3_{c}}_{s=1}(cc)^{I=0,\bar{3}_{c}}_{s=1}\rangle^{I=1}_{s=1}\\
|[(ns)^{I=\frac{1}{2},\bar{3}_{c}}_{s=0}(ns)^{I=\frac{1}{2},\bar{3}_{c}}_{s=1}]^{I=1,\bar{6}_{c}}_{s=1}(cc)^{I=0,6_{c}}_{s=0}\rangle^{I=1}_{s=1}
\end{pmatrix}$}
&
\multirow{2}*{$\begin{pmatrix}
5696\\5608
\end{pmatrix}$}
&1377.8&-1601.8&0.6&1.27&1.25&1.36&1.33&1.42&1.26&\multicolumn{2}{c}{14.0}&\multicolumn{2}{c|}{13.6}&27.6
\\
&&&1366.5&-1683.7&6.2&1.47&1.46&1.33&1.27&1.11&1.46&
\multicolumn{2}{c}{9.6}&\multicolumn{2}{c|}{9.2}&18.8
\\
\toprule[0.10pt]
\multirow{1}*{$1(0)[0^{+}]$}&$|[(ns)^{I=\frac{1}{2},\bar{3}_{c}}_{s=0}(ns)^{I=\frac{1}{2},\bar{3}_{c}}_{s=1}]^{I=1,3_{c}}_{s=1}(cc)^{I=0,\bar{3}_{c}}_{s=1}\rangle^{I=1}_{s=0}$&
5691&1380.3&1604.3&-4.0&1.27&1.25&1.36&1.32&1.42&1.26&\multicolumn{2}{c}{}&\multicolumn{2}{c|}{13.3}&13.3
\\
\bottomrule[0.50pt]
\bottomrule[1.50pt]
\end{tabular}
\end{lrbox}\scalebox{0.81}{\usebox{\tablebox}}
\end{table*}

Finally, we discuss the $cc[nn][ss]$ and $cc[ns][ns]$ subsystems.
These two subsystems both have the same quark contents, same quantum numbers, and same mass range.
However, we can still distinguish them easily,
because their decay final states show obvious differences. 
The $cc[nn][ss]$ states mainly decay to $\Sigma^{(*)}_{c}\Omega^{(*)}_{c}$ and $\Lambda_{c}\Omega^{(*)}_{c}$ final states,
with the $\Xi^{(*)}_{c}\Xi^{(*)}_{c}$ decay channel being extremely suppressed.
In contrast, the decay behavior of the $cc[ns][ns]$ states is completely opposite to that of the $cc[nn][ss]$ states.

For isovector $cc[nn][ss]$ states, they are all relatively broad states, with their total widths ranging from 30 to 70 MeV.
Then for the four $I[J^{P}]=1[1^{+}]$ states, three of them, namely ${\rm H}_{c^{2}[nn][ss]}(5811, 1, 1^{+})$, ${\rm H}_{c^{2}[nn][ss]}(5810, 1, 1^{+})$, and ${\rm H}_{c^{2}[nn][ss]}(5809, 1, 1^{+})$, are partner states with similar masses and widths.
Their total widths are 60, 69, and 64 MeV, respectively, and the mass gaps among them are only 1 MeV.
We can distinguish them by the ratios of relative partial widths:
\begin{eqnarray}\label{width3}
\Gamma_{\Sigma^{*}_{c}\Omega^{*}_{c}}:\Gamma_{\Sigma^{*}_{c}\Omega_{c}}:\Gamma_{\Sigma_{c}\Omega^{*}_{c}}:\Gamma_{\Sigma_{c}\Omega_{c}}=6:1:14:33, \nonumber\\
\Gamma_{\Sigma^{*}_{c}\Omega^{*}_{c}}:\Gamma_{\Sigma^{*}_{c}\Omega_{c}}:\Gamma_{\Sigma_{c}\Omega^{*}_{c}}:\Gamma_{\Sigma_{c}\Omega_{c}}=1:2:23:0.3, \nonumber
\end{eqnarray}
and
\begin{eqnarray}\label{width4}
\Gamma_{\Sigma^{*}_{c}\Omega^{*}_{c}}:\Gamma_{\Sigma^{*}_{c}\Omega_{c}}:\Gamma_{\Sigma_{c}\Omega^{*}_{c}}:\Gamma_{\Sigma_{c}\Omega_{c}}=22:1:35:1,
\end{eqnarray}
respectively.
From the above ratios, we notice that $\Sigma_{c}\Omega_{c}$ channel is the dominant decay channel for the ${\rm H}_{c^{2}[nn][ss]}(5811, 1, 1^{+})$.
In contrast, $\Sigma_{c}\Omega_{c}$ channel is suppressed in the ${\rm H}_{c^{2}[nn][ss]}(5810,1, 1^{+})$ and ${\rm H}_{c^{2}[nn][ss]}(5809, 1, 1^{+})$.
They mainly decay to $\Sigma_{c}\Omega^{*}_{c}$ final states and $\Sigma_{c}\Omega^{*}_{c}$, $\Sigma^{*}_{c}\Omega^{*}_{c}$ final states, respectively.

For isoscalar states, there exists a narrow state, ${\rm H}_{c^{2}[nn][ss]}(5586, 0, 0^{+})$. It has a total width of about 7 MeV and only decays to $\Lambda_{c}\Omega_{c}$ final states.
Although it has a larger decay phase space, the signs of the Feynman amplitudes $\mathcal{M}(A\to BC)$ (Eq.~(\ref{amp})) from the four quark-interchange diagrams (Fig. \ref{fig3}) are different for the $\Lambda_{c}\Omega_{c}$ decay channel. 
The contributions among them largely cancel out, leading to the suppression of the decay width.
These characteristics, namely the narrow width and the unique $\Lambda_{c}\Omega_{c}$-only decay mode, are highly desirable in experimental searches. 
They make the lineshape of ${\rm H}_{c^{2}[nn][ss]}(5586, 0, 0^{+})$ more prominent in relevant experiments, thus increasing the likelihood of its discovery. 
Therefore, we suggest that experiments prioritize the search for possible resonance peaks in the $5500$-$5600$ MeV range of the $\Lambda_{c}\Omega_{c}$ invariant mass spectrum. 

For the $cc[ns][ns]$ subsystem, the two $[ns]$ diquarks need to satisfy the Spin-Statistics theorem as identical particles in the $|[(ns)^{I=1/2}_{s=1}(ns)^{I=1/2}_{s=1}](cc)\rangle$ and $|[(ns)^{I=1/2}_{s=0}(ns)^{I=1/2}_{s=0}](cc)\rangle$ configurations.
In the $|[(ns)^{I=1/2}_{s=0}(ns)^{I=1/2}_{s=1}](cc)\rangle$ configuration, since the two diquarks are not identical particles, the constraints imposed on the 
color-spin wave functions for their coupling to the $I=1$ state and the $I=0$ state are identical.
As a result, they have exactly the same masses, RMS radii, and decay behaviors.
Among them,  ${\rm H}_{c^{2}[ns]^{2}}(5828, 1, 1^{+})$ is the state with the narrowest width.
Although it has multiple rearrangement decay channels: $\Xi^{*}_{c}\Xi^{*}_{c}$, $\Xi^{*}_{c}\Xi'_{c}$, and $\Xi'_{c}\Xi'_{c}$, its total width is still less than 10 MeV.
The partial width ratio is:
\begin{eqnarray}\label{width5}
\Gamma_{\Xi^{*}_{c}\Xi^{*}_{c}}:\Gamma_{\Xi^{*}_{c}\Xi'_{c}}:\Gamma_{\Xi'_{c}\Xi'_{c}}=1:2.5:0.3.
\end{eqnarray}
Our results show that $\Xi^{*}_{c}\Xi'_{c}$ is the dominant decay channel.
We propose that future experimental investigations could explore the presence of the signal corresponding to ${\rm H}_{c^{2}[ns]^{2}}(5828, 1, 1^{+})$ within the mass range of 5800-5900 MeV in the $\Xi^{*}_{c}\Xi'_{c}$ final states.
For other $cc[ns][ns]$ states, one can perform similar discussions on the decay behaviors according to Table \ref{ccnnss} and Fig.~\ref{fig-ccnnss}.
\\

\subsection{Discussions of the coupled-channel effects}

In the above studies, we mainly focus on the quenched situations. According to Figs.~\ref{fig-ccnnnn}-\ref{fig-ccnnss}, some states have widths with tens of MeV. In this way, the coupling between the original states and channels may change the masses, i.e., the coupled-channel effects~ \cite{Li:2012cs,Duan:2020tsx,Luo:2019qkm,Duan:2021bna,Duan:2021alw,Man:2025zfu}. In addition, some of the same $J^P$ may contain several states, which leads to more complex coupled-channel effects, i.e., the different states with the $J^P$ quantum numbers could be coupled by the common channels. 
In order to quantitatively scale how large the mass shifts from coupled-channel effects, we introduce the following coupled-channel equation, i.e.~\cite{Man:2025zfu,Man:2025vmm},
\begin{equation}\label{eq:cpe}
\begin{split}
&\left(\begin{array}{cc}
M_{\rm bare}^1 + \Delta M_{11}(M) & \Delta M_{12}(M)\\
\Delta M_{21}(M)     & M_{\rm bare}^2 + \Delta M_{22}(M)
\end{array}\right)\left(\begin{array}{c}
c_1\\c_2
\end{array}\right)\\
&=M\left(\begin{array}{c}
c_1\\c_2
\end{array}\right),
\end{split}
\end{equation}
where $M$ is the eigenvalue of the equation, which is also the physical mass obtained by the coupled-channel effects. Different from the physical mass $M$, $M_{\rm bare}^1$ and $M_{\rm bare}^2$ are bare masses of two states, i.e., the results from pure potential model. The $\Delta M_{11}$, $\Delta M_{22}$, $\Delta M_{12}$, and $\Delta M_{21}$ are mass shifts, which are defined by
\begin{equation}\label{eq:M11}
\Delta M_{11}(M)=\frac{1}{2J_{A_1}+1}\sum\limits_{B_iC_i}{\rm Re}\int \frac{|T(A_1\to B_iC_i)|^2}{M-E_{B_i}-E_{C_i}}{\rm d}^3{\bf P}_{B_i},
\end{equation}
\begin{equation}\label{eq:M22}
\Delta M_{22}(M)=\frac{1}{2J_{A_2}+1}\sum\limits_{B_iC_i}{\rm Re}\int \frac{|T(A_2\to B_iC_i)|^2}{M-E_{B_i}-E_{C_i}}{\rm d}^3{\bf P}_{B_i},
\end{equation}
\begin{equation}\label{eq:M12}
\begin{split}
&\Delta M_{12}(M)=\frac{1}{2J_{A_1}+1}\\
&\quad\sum\limits_{B_iC_i}{\rm Re}\int \frac{T(A_1\to B_iC_i)T^*(A_2\to B_iC_i)}{M-E_{B_i}-E_{C_i}}{\rm d}^3{\bf P}_{B_i},
\end{split}
\end{equation}
\begin{equation}\label{eq:M21}
\begin{split}
&\Delta M_{21}(M)=\frac{1}{2J_{A_1}+1}\\
&\quad\sum\limits_{B_iC_i}{\rm Re}\int \frac{T^*(A_1\to B_iC_i)T(A_2\to B_iC_i)}{M-E_{B_i}-E_{C_i}}{\rm d}^3{\bf P}_{B_i},
\end{split}
\end{equation}
where $A_{1(2)}$ and $B_iC_i$ are bare states and intermediate channels, respectively. 
$T(A_{1(2)}\to B_iC_i)$ is the transition matrix element of $A_{1(2)}\to B_iC_i$, which is defined in Eqs.~(\ref{Eq:T1})-(\ref{Eq:T2}). 
In this work, we take the $H_{c^2[ss]^2}(5950,0^+)$ and $H_{c^2[ss]^2}(6025,0^+)$ as examples to illustrate the contributions
from coupled-channel effects. 
In this scheme, $A_1$ and $A_2$ correspond to $H_{c^2[ss]^2}(5950,0^+)$ and $H_{c^2[ss]^2}(6025,0^+)$, respectively. 
As shown in Fig.~\ref{fig-ccnnnn}, the two $0^+$ states could decay into $\Omega_c\Omega_c$ and $\Omega_c^*\Omega_c^*$ in $S$-wave. 
Thus, we take the intermediate channels $B_iC_i$ as $\Omega_c\Omega_c$ and $\Omega_c^*\Omega_c^*$. 
Finally, we obtain two solutions, i.e., 5959 MeV and 6039 MeV. We find the mass shift is only about 10 MeV and the sign is positive. We notice that even though the two $0^+$ states could couple with $\Omega_c\Omega_c$ and $\Omega_c^*\Omega_c^*$ with $S$-wave, the bare masses are about 400-600 MeV higher than the thresholds. In this work, we find that the coupled-channel effects are significant near the thresholds, i.e., the mass shifts may be larger than 40 MeV. This property matches the results in Refs.~\cite{Luo:2019qkm,Duan:2020tsx}. But in this work, the bare masses are much higher than the thresholds, which leads to the coupled-channel effects being highly depressed. On the other hand, we notice that the coupled-channel effects enlarge the masses. If the masses $M$ are below the thresholds, Eqs.~(\ref{eq:M11})-(\ref{eq:M22}) are general integrations, and the signs are negative. But if the masses $M$ are above the thresholds, we should take the principal values of integrations in Eqs.~(\ref{eq:M11})-(\ref{eq:M21}), and the signs are not determined. In some situations, the coupled-channel effects may increase the masses~\cite{Duan:2020tsx}.

\section{summary}\label{sec4}

In the field of hadron physics, the study of novel exotic states is a central research focus.
Meanwhile, the diquark-heavy quark picture has been widely applied to singly charmed baryon systems, yielding theoretical results that are in high agreement with experimental outcomes.  
Inspired by this success and based on the observed $T_{cc}^+(3875)$, this study further extends the investigation by replacing the two antiquarks $\bar{q}$ with two strongly correlated light diquarks $[qq]$, forming the new doubly charmed hexaquark system ($cc[qq][qq]$).  
An in-depth study is conducted within the diquark-diquark-heavy quark-heavy quark picture framework, aiming to provide a more realistic mass spectrum for the doubly charmed hexaquark system.


In this research, we first construct total wave functions that satisfy the 
Spin-Statistics Theorem, covering the flavor, spatial, color, and spin parts. 
Then, within the framework of the constituent quark model, we systematically calculated the mass spectra of the 
doubly charmed hexaquark system using the Gaussian expansion method.
Finally, through detailed calculations, we also obtained corresponding internal mass contributions, root-mean-square radii, two-body strong decay partial widths, and total decay widths.
In addition, we conducted a dedicated discussion of coupled-channel effects. 

The mass spectra of the doubly charmed hexaquark system are in the range of $5000$–$6000$ MeV, and the analysis of the mass spectra shows that there is no stable state in this system.
All states can decay into two final-state singly-charmed baryons through two-body rearrangement strong decay. 
The analysis of the internal mass contributions shows that the contributions of the kinetic energy and the confinement potential are of the same order of magnitude, 
while the contribution of the hyperfine interaction potential is suppressed and relatively small, leading to low configuration mixing and small mass gaps between certain configurations, resulting in partner states.

Regarding the RMS radii, 
most results are in the range of $1.2$-$1.6$ fm, 
roughly in the same order of magnitude.
This implies that the spatial distribution between quarks is relatively close and the internal interactions within the system are strong, which is consistent with the expectations of the compact doubly charmed hexaquark configuration.

After that, we deeply analyzed the decay behavior of the doubly charmed hexaquark system.
The analysis indicates that most observed states possess total decay widths ranging from 30 to 100 MeV.
However, there are still some narrow states. 
Even though these states generally have a large two-body decay phase space, the total widths of some of them are even less than 10 MeV.
Upon further investigation of the reason, we found that for some specific decay channels, the signs of the Feynman amplitudes $\mathcal{M}(A\rightarrow BC)$ from the four quark-exchange diagrams are opposite. This causes their contributions to largely cancel each other out, ultimately leading to the suppression of the decay width. 

In the doubly charmed hexaquark system, 
the ${\rm H}_{c^{2}[nn]^{2}}(5043,0,0^{+})$ state is an ideal candidate for experimental detection. 
Composed of two scalar $[nn]$ diquarks ($I = 0, S = 0$), its structure enhances the internal confinement potential and chromomagnetic interaction, 
resulting in the lowest mass (5043 MeV) in the mass spectrum. 
It mainly decays to the $\Lambda_{c}\Lambda_{c}$ final state with a total width of 14.3 MeV, with a narrow hadronic resonance characteristic.
Given the experimental detectability of doubly charmed multi-quark systems confirmed by the discovery of \(T_{cc}^+(3875)\), we suggest the LHCb, CMS, ATLAS collaborations to analyze the $\Lambda_{c}\Lambda_{c}$ invariant mass spectrum in the $5000$–$5100$ MeV range in high-luminosity $pp$ collisions to search for this state. 

Furthermore, several narrow states also show great potential for experimental detection. 
The ${\rm H}_{c^{2}[nn][ns]}(5482, 1/2, 0^{+})$ state in the $cc[nn][ns]$ subsystem has a total width of less than 5 MeV and decays only to $\Lambda_{c}\Xi'_{c}$, with high detectability in the $5400$-$5500$ MeV window.
The ${\rm H}_{c^{2}[nn][ss]}(5586, 0, 0^{+})$ state in the $cc[nn][ss]$ subsystem has a total width of about 7 MeV and decays only to $\Lambda_{c}\Omega_{c}$. 
It may form a resonance peak in the 5500-5600 MeV range of the $\Lambda_{c}\Omega_{c}$ invariant mass spectrum. 
The ${\rm H}_{c^{2}[ns]^{2}}(5828, 1, 1^{+})$ state in the $cc[ns][ns]$ subsystem has a total width of less than 10 MeV and its dominant decay channel is $\Xi_{c}^{*} \Xi_{c}'$, showing distinct lineshape compared to the background in the $5800$-$5900$ MeV range.

In addition, we also take $H_{c^2[ss]^2}(5950,0^+)$ and $H_{c^2[ss]^2}(6025,0^+)$ as examples to illustrate the contributions from the coupled-channel effects. Since the masses are much higher than the thresholds, the coupled-channel effects are very small.

In summary, these results comprehensively reveal the mass spectra, internal structures, and decay characteristics of doubly charmed hexaquark states.
We hope they can provide some perspectives for further theoretical research. 
Meanwhile, we also look forward to more experimental collaborations focusing on doubly charmed hexaquark states in the future. 
Conducting more experimental measurements can not only test our research results, but also deepen the understanding of the interactions within multi-quark systems.

\section*{Acknowledgements}

H.-T. An is supported by the National Nature Science Foundation of China under Grant No.12447172, by the Postdoctoral Fellowship Program of CPSF under Grant No.GZC20240877, and by Shuimu Tsinghua Scholar Program of Tsinghua University under Grant No.2024SM119. This work is also supported by the National Natural Science Foundation of China under Grant Nos. 12335001, 12247101, and 12405098, the ``111 Center" under Grant No. B20063, the Natural Science Foundation of Gansu Province (No. 22JR5RA389 and No. 22JR5RA171), the fundamental Research Funds for the Central Universities (Grant No. lzujbky-2023-stlt01), and the project for top-notch innovative talents of Gansu province.

\bibliographystyle{UserDefined}
\bibliography{References}

\end{document}